\documentclass[
aps,
pra,
superscriptaddress,
amssymb,
reprint,
]{revtex4-1}

\usepackage[letterpaper, margin=0.75in]{geometry}

\usepackage{amsmath}
\usepackage{amssymb}
\usepackage{outlines}
\usepackage{comment}
\usepackage{graphicx}
\usepackage{siunitx}
\usepackage{physics}
\usepackage{setspace}

\newcommand{\mytitle} {Error-detected state transfer and entanglement in a superconducting quantum network}

\graphicspath{{../../figures/v6/}}
\newcommand{\md} {\textemdash{} }
\newcommand{\supp} {(Supplementary Information)}
\newcommand{\fig}[1] {Fig. #1}
\newcommand{\figref} {Fig. }
\newcommand{\figsref} {Figs. }
\newcommand{\eqn} {equation }
\newcommand{\eqnref}[1] {\eqn(\ref{#1})}
\newcommand{\tabref} {Table }
\newcommand{\refref} {Ref. }
\newcommand{\secref} {Sec. }
\newcommand{\ssecref} {Subsec. }

\newcommand{\modeb} {\hat{b}}
\newcommand{\modea}[1] {\hat{a}_{#1}}

\renewcommand{\Tr}[1] {\mathrm{Tr}\left( #1 \right)}

\newcommand{\chiaa} {\chi_\mathrm{aa}}
\newcommand{\chitt} {\chi_\mathrm{tt}}
\newcommand{\chicc} {\chi_\mathrm{cc}}
\newcommand{\chiac} {\chi_\mathrm{ac}}
\newcommand{\chiat} {\chi_\mathrm{at}}
\newcommand{\chibc} {\chi_\mathrm{bc}}
\newcommand{\chirt} {\chi_\mathrm{rt}}

\newcommand{\tbs} {\tau_\mathrm{BS}}
\newcommand{\tswap} {\tau_\mathrm{SWAP}}
\newcommand{\tfty} {\tau_\mathrm{50:50}}
\newcommand{\kappab} {\kappa_\mathrm{b}}

\newcommand{\ketL}[1] { \ket{ #1 }_\mathrm{L} }
\newcommand{\ketE}[1] { \ket{ #1 }_\mathrm{E} }
\newcommand{\rhoL} { \rho_\mathrm{L} }

\newcommand{\yaleaffil}
{
\affiliation{Departments of Applied Physics and Physics, Yale University, New Haven, CT 06520, USA
}
\affiliation{Yale Quantum Institute, New Haven, CT 06520, USA
}
}

\graphicspath{{./figures/},{./supplement_figs/}}

\begin{document}

\begin{abstract}\textbf{
Modular networks are a promising paradigm for increasingly complex quantum devices based on the ability to transfer qubits and generate entanglement between modules \cite{Kimble2008, Gottesman1999, Jiang2007a}.
These tasks require a low-loss, high-speed intermodule link that enables extensible network connectivity.
Satisfying these demands simultaneously remains an outstanding goal for long-range optical quantum networks \cite{Kimble2008,Northup2014} as well as modular superconducting processors within a single cryostat.
We demonstrate communication and entanglement in a superconducting network with a microwave-actuated beamsplitter transformation between two bosonic qubits, which are housed in separate modules and joined by a demountable coaxial bus resonator.
We transfer a qubit in a multi-photon encoding \cite{Leghtas2013} and track photon loss events to improve the fidelity, making it as high as in a single-photon encoding.
Furthermore, generating entanglement with two-photon interference and postselection against loss errors produces a Bell state with success probability 79\% and fidelity 0.94, halving the error obtained with a single photon.
These capabilities demonstrate several promising methods for faithful operations between modules, including novel possibilities for resource-efficient direct gates \cite{Lau2016, Gao2019}.
}
\end{abstract}

\title{\mytitle}

\author{
Luke D. Burkhart}
\email{luke.burkhart@yale.edu}
\author{
James Teoh}
\author{
Yaxing Zhang}
\yaleaffil
\author{
Christopher J. Axline}
\yaleaffil
\affiliation{Present address: Institute for Quantum Electronics, ETH Z\"urich, Otto-Stern-Weg 1, 8093 Z\"urich, Switzerland
}
\author{
Luigi Frunzio}
\author{
M.H. Devoret}
\yaleaffil
\author{
Liang Jiang}
\yaleaffil
\affiliation{Pritzker School of Molecular Engineering, The University of Chicago, Illinois 60637, USA}
\author{
S.M. Girvin}
\author{
R.J. Schoelkopf}
\email{robert.schoelkopf@yale.edu}
\yaleaffil
\date{\today}
\maketitle

\begin{figure}[t]
    \input{fig_text/figure1}
\end{figure}

\begin{figure}[t]
    \input{fig_text/figure2}
\end{figure}

\begin{figure*}[t]
    \input{fig_text/figure3}
\end{figure*}

\begin{figure*}[t]
    \input{fig_text/figure4}
\end{figure*}

Quantum networks must address the problem of inefficiencies in signal coupling, transmission, routing, and receiving, both by minimizing losses in these stages and by mitigating the impact of loss on operation fidelities.
Both optical fiber and microwave transmission line can be very efficient transmission channels, with loss over a meter of superconducting coaxial cable as low as $10^{-3}$ \cite{Kurpiers2017a}.
Circuit quantum electrodynamics (cQED) combines this low transmission loss with efficient switchable couplings between qubits and microwave photons \cite{Pechal2014,Narla2016, Pfaff2017}, making it a promising candidate for modular quantum computing.
However, robust and scalable networks require communication protocols with performance beyond even small losses.
Schemes which condition on success with an ancillary herald are employed in both optical \cite{Northup2014} and microwave networks \cite{Narla2016,Kurpiers2018a}, at the cost of reduced operation rate.
The cQED toolkit also enables quantum non-demolition (QND) measurement of multi-photon bosonic qubits for deterministic correction of loss \cite{Ofek2016}, though realization of this approach for communication has remained an outstanding challenge \cite{Michael2016,Axline2018}.

cQED networks that employ propagating photons \cite{Narla2016, Roch2014, Dickel2018, Kurpiers2018, Axline2018, Campagne-Ibarcq2018} have been unable to leverage the intrinsic quality of superconducting cable due to reliance on lossy directional elements such as circulators.
In contrast, intermodule links that forgo directional elements can have much lower loss, and support a standing-mode structure \cite{Leung2019, Zhong2019}.
A single mode can be used as a quantum bus \cite{Majer2007, Sillanpaa2007}, potentially connecting several modules (\fig{1a}).
This simple link provides features unavailable in a directional channel, such as interference of photons from different modules.
When the bus connects two bosonic modes $\modea{1,2}$, this interference can give rise to an effective beamsplitter transformation:
\begin{equation}\label{eq:beamsplitter}
\begin{aligned}
\modea{1} &\rightarrow \cos\theta \,\,\modea{1} + \sin\theta \,\,\modea{2} \\
\modea{2} &\rightarrow \cos\theta \,\,\modea{2} - \sin\theta \,\,\modea{1} .
\end{aligned}
\end{equation}
The beamsplitter is a powerful tool for manipulating and entangling propagating photons \cite{Kok2007,Aaronson2013}.
In this work we extend these applications to stationary modes in separate modules, using this interaction for two key networking tasks \textemdash{} qubit transfer and entanglement generation.

As a simple demonstration of the quantum network bus, we connect two modules with a superconducting coaxial cable of length $l=\SI{6.6}{\cm}$, though this technique extends to longer links \supp.
We choose the third harmonic ($l=3\lambda/2$) of the cable as the bus, shown in \fig{1b}.
Constructed without lossy microwave components or connectors, this mode has a relatively high quality factor ($Q=51,000$, $\kappab/2\pi = \SI{110}{\kHz}$).
Each module contains a 3D cavity bosonic qubit coupled to the bus via a conversion transmon \cite{Axline2016}.
When the transmon is driven by two off-resonant pumps with frequency difference close to the difference between the cavity ($\modea{j}$) and bus ($\modeb$) frequencies, the Josephson nonlinearity enables parametric conversion \cite{Pfaff2017, Zhang2019} of the form
\begin{equation}\label{eq:H_pc}
H_j/\hbar = ge^{-i\Delta_j t}\modea{j}^\dag \modeb + g^*e^{i\Delta_j t} \modea{j} \modeb^\dag .
\end{equation}
The pump amplitudes control the conversion rate $g$, and their frequencies set the detuning $\Delta_j$ from the frequency-matching condition.

An important consideration when utilizing this coupling is to ensure that no information is left in the bus at the end of the protocol.
When conversion is activated in both modules simultaneously, the resulting three-mode system allows bidirectional population transfer in which the bus begins and ends in the vacuum state (\fig{1c,d}), satisfying this requirement.
However, constant resonant conversion ($\Delta_1=\Delta_2=0$) is not well suited for entanglement by partial transfer, because halfway through transfer the occupation of the bus is maximized, leaving it entangled with the cavities.
There are solutions to this problem that involve modulating the coupling strengths in time, such as stimulated Raman by adiabatic passage \cite{Bergmann1998}, but this approach does not allow the full range of interference effects utilized here \supp.
Instead, we choose to use equal, nonzero detuning in both modules ($\Delta_1=\Delta_2=\Delta$).
The best-known case is the virtual Raman regime $\Delta \gg g$, where the bus occupation is suppressed, effectively eliminating it from the dynamics, resulting in the beamsplitter evolution of \eqnref{eq:beamsplitter} between the cavities.
This elimination suppresses infidelity due to loss in the bus, but lengthens the protocol significantly, an unfavorable trade-off when other sources of decoherence demand faster operations.

We address these concerns by working at intermediate detuning $\Delta \sim g$, where the beamsplitter transformation in \eqnref{eq:beamsplitter} can be constructed with any mixing angle $\theta$ without slowing down the protocol.
We take the conversion rate as fixed, since it is usually limited by experimental constraints, and use detuning as a control parameter.
For a given detuning, the bus returns to vacuum and is eliminated periodically, with period $\tbs = 2\pi/\sqrt{8g^2 + \Delta^2}$.
Thus the evolution is a linear transformation on the two cavities of the form of \eqnref{eq:beamsplitter}, with beamsplitter angle
\begin{equation}
\theta(\Delta) = \frac{\pi}{2} \left( 1 - \frac{ \Delta \: \tbs}{2\pi} \right)
\end{equation}
\supp.
This operation, tunable from opaque to transparent, is equivalent to a physical beamsplitter for propagating photons as depicted in \fig{2a}, but we emphasize that here this evolution applies to two stationary modes, whose initial (final) states are represented by the input to (output from) the beamsplitter.
By relying on parametric conversion and a high-quality bus, this scheme has the benefit compared to previous experiments \cite{Narla2016,Roch2014,Dickel2018,Axline2018,Campagne-Ibarcq2018,Kurpiers2018,Leung2019,Zhong2019} of not requiring precise frequency matching or tunability.

We apply this idea to eliminate the bus at two beamsplitter working points \textemdash{} fully transparent for state transfer, and 50:50 for entanglement generation.
We prepare a single photon in cavity 1, turn on conversion in both modules, and measure the occupation of each cavity as a function of time, shown in \fig{2b}, for $g/2\pi=560$ kHz, the value used throughout this work.
At zero detuning ($\theta = 90^\circ$), we observe complete excitation transfer between cavities in $\tswap =624$ ns, with energy transfer efficiency $\eta = 84\%$.
Furthermore, an intermediate detuning $\Delta=\sqrt{8/3}g$ realizes a 50:50 beamsplitter ($\theta=45^\circ$), which can generate entanglement from separable input states.
The time to eliminate the bus at this detuning is $\tfty=520$ ns, demonstrating the advantage of this operating point over the much slower virtual Raman regime.
The dynamics as a function of detuning in \fig{2c} demonstrate the continuous tunability of this operation.

Both operations are several hundred times faster than the decay rate of either bosonic qubit ($\kappa_{1,2}^{-1}=300, 450$ \si{\micro\second}), so the excitation decay during the beamsplitter is dominated by dissipation in the bus ($\kappab^{-1} = 1.5$ \si{\micro\second}).
The transfer inefficiency due to loss is $1{-}\eta \approx \kappab/2g = 11\%$; thus, these operations can be improved in future with increased conversion rate or bus quality, which is partly limited by resistive loss at the interface with the modules \supp.
Additionally, excitation events of either conversion transmon due to thermal or pump-induced transitions \cite{Zhang2019} dispersively shift the bus and cavity frequencies, effectively turning off conversion and dephasing the cavity.
These events are responsible for an additional 3\textendash 4\% inefficiency, but could be mitigated with a different type of conversion element \cite{Frattini2017}.
The value of $g$ used here is a trade-off between speed, which reduces loss errors, and excitation errors induced by the pumps; an element which makes this trade-off more favorable can in future improve the beamsplitter.

To characterize the bus as a communication link, we turn to the task of state transfer, an operation which can move qubits between modules for local processing or entanglement distribution.
We demonstrate transfer of a single qubit from one module to another; however, the link is bidirectional and linear, allowing simultaneous two-way transfer for any input states.
The simplest encoding is the Fock basis \textemdash{} presence or absence of a single photon.
We prepare a known superposition $\alpha |0\rangle + \beta |1\rangle$ in cavity 1, apply the $90^\circ$ beamsplitter, and perform Wigner tomography on the resulting state in cavity 2 (\fig{3a}).
This protocol is performed for six input states, two of which are shown in \fig{3b}.
The resulting mean state fidelity, reconstructed using maximal likelihood estimation \supp, is $\bar{F}_\text{Fock} = 0.92(1)$, the average fidelity with which an arbitrary quantum state is transfered.

This protocol yields significant improvement over previous experiments which relied on directional communication links \cite{Axline2018,Kurpiers2018}, but the performance remains dominated by loss in the bus.
We demonstrate suppression of this source of infidelity by leveraging strategies for correcting photon loss in stationary memories \cite{Leghtas2013,Michael2016,Ofek2016}.
Because the cavities and the bus are all harmonic oscillators, we use an encoding which detects loss in any of these modes with a single measurement.
We choose the cat code \cite{Leghtas2013}, superpositions of four coherent states with even photon number parity, and add a QND parity measurement using an ancilla transmon prior to Wigner tomography (\fig{3c}).
The cat code employed here is chosen to contain on average $\bar{n} \approx 1.7$ photons, more than three times as many as the Fock encoding.
This increase makes loss events more likely, and is the unavoidable overhead of an error-correctable encoding.
Indeed, without error detection, the mean state fidelity is $\bar{F}_\text{cat} = 0.80(1)<\bar{F}_\text{Fock}$ \supp.
However, the parity measurement can detect loss events, mitigating the impact on fidelity and resulting in a net gain.

By adding the syndrome measurement before tomography, we overcome the overhead to reach the break-even point with respect to the Fock encoding.
The measured Wigner functions sorted by parity outcome are shown in \fig{3d}.
The dominant outcome is that the parity remains even ($p_\text{even}=84\%$), and the fidelity in this no-error case is $\bar{F}_\text{even} = 0.93(1)$.
Change in parity denotes a single loss error ($p_\text{odd}=16\%$), and we find the coherence to be preserved in the odd manifold, with fidelity $\bar{F}_\text{odd} = 0.86(3)$.
This error can be corrected in real time, or the state conditionally decoded \cite{Ofek2016} \supp, but the odd manifold is also a valid logical encoding, so future operations may simply be updated without feedback correction; hence, the state is error-tracked.
Averaged over all trials, the resulting deterministic fidelity is $\bar{F}_\text{cat,tracked} = 0.92(2)$, which includes encoding and syndrome measurement errors of $\sim 3\%$ \supp.
Even in the presence of code overhead and these imperfections, the cat code still reaches the break-even point at which the transfer fidelity is as high as in the Fock encoding.
In future, use of a more robust parity measurement \cite{Rosenblum2018} and suppression of conversion transmon excitations \textemdash{} an uncorrectable error \textemdash{} should enable error-corrected state transfer beyond break-even.

To fulfill another key network requirement, we generate entangled states between modules, a critical resource for non-local gates \cite{Gottesman1999,Chou2018,Wan2019} and stabilizer measurements \cite{Nickerson2013}.
The 50:50 beamsplitter in \fig{2b} can entangle the modules for a variety of input states.
The simplest input is a single photon in one of the cavities ($|10\rangle$), ideally creating the odd-parity Bell state $\left(|10\rangle + |01\rangle\right)/\sqrt{2}$ (\fig{4a}).
To characterize the entanglement, we perform logical two-qubit tomography by measuring conditional Wigner functions.
We measure cavity 2 in one of three logical bases \{X, Y, Z\} \supp, ideally projecting cavity 1 into one of the corresponding basis states depending on the measurement result.
The Wigner function of cavity 1, measured simultaneously and conditioned on the logical measurement outcome, shows the correlations expected of an entangled state (\fig{4b}).
From these data we reconstruct the logical two-qubit state \supp, which has fidelity to the ideal Bell state $\bar{F}_\text{Bell,01} = 0.88(1)$, not corrected for a $2\%$ error in the logical measurement \supp.
The reconstructed joint Pauli expectation values (\fig{4c}) exhibit dominant two-qubit correlations, but visible single-qubit polarization indicates the infidelity results mostly from photon loss in the bus.

Finally, we implement a novel scheme for QND detection of loss during entanglement generation, which is probabilistic but results in higher fidelity when successful.
We apply the 50:50 beamsplitter to the two-photon input state $|11\rangle$, ideally producing $\left(|20\rangle + |02\rangle\right)/\sqrt{2}$, an example of Hong\textendash{}Ou\textendash{}Mandel interference \cite{Hong1987} that relies on frequency conversion to make photons indistinguishable.
Single-photon loss results in odd photon number occupation in one of the cavities after the beamsplitter, and can be detected by measuring the parity of each cavity, as indicated in \fig{4d}.
Since the ideal state has only even number, these measurements do not dephase the entangled state.
We declare successful entanglement when both cavities have even parity (success probability 79\%), and post-select the resulting data.
The measured conditional Wigner functions in \fig{4e} exhibit the expected correlations in the two-photon manifold, and the reconstructed logical state in \fig{4f} shows reduced single-qubit polarization compared to the single-photon protocol.
The Bell state fidelity, again uncorrected for tomography, is $\bar{F}_\text{Bell,02} = 0.94(1)$, a two-fold reduction of errors from the single-photon case, with only a small failure rate, which can be improved by reducing loss in the bus.
We emphasize that this simple protocol is only possible with bosonic qubits and a bidirectional bus which supports multiple indistinguishable photons.

This error-detected protocol also admits a simple and rapid way to increase the effective success probability.
Because the state upon failure is known ($|01\rangle$ or $|10\rangle$), we can simply load another photon into the empty cavity and reapply the beamsplitter.
For instance, with up to three entangling attempts, the success probability increases to 95\%, with only a small decrease in the fidelity to $\bar{F}_\text{Bell,02} = 0.91(1)$, with an average entangling time of 5.2 \si{\micro\second}, twenty times faster than the cavity decoherence times \supp.
This ability to rapidly generate high-fidelity Bell states is an essential feature for teleportation schemes using resource entanglement.

We have demonstrated a high-quality quantum bus in a cQED network which enables a flexible beamsplitter operation between bosonic qubits in separate modules.
We use the beamsplitter to transfer qubits and generate entanglement, and show how both protocols can be improved with multi-photon states and error-detection, reaching the break-even point for state transfer, and reducing the Bell state infidelity by half with a high success probability.
The bus efficiency can be significantly enhanced with reasonable improvements in assembly, a higher-quality transmission line or waveguide link, or improved parametric conversion.
This platform offers routes towards extensible networks by coupling several modules to a single bus, or arraying modules with multiple two-port buses.

The capabilities demonstrated here suggest several complementary approaches for operations between modules in superconducting quantum networks.
The entanglement fidelity achieved completes the set of requirements for a viable gate over the network \cite{Chou2018}, while error-correctable state transfer enables shuttling logical qubits between modules to instead perform local gates.
The low-loss beamsplitter also allows application of gates such as controlled-SWAP and exponential-SWAP \cite{Lau2016, Gao2019} directly across the network without resource entanglement or shuttling.
Finally, the conversion interaction employed here is a general tool for manipulating bosonic degrees of freedom in separate modules, which can be applied for boson sampling \cite{Aaronson2013,Wang2019} and as a tunable hopping term for quantum simulation of bosons.
This diverse set of possibilities in a single platform provides many directions for future research towards distributed quantum computing with superconducting networks.

%

\section*{Acknowledgments}
The authors acknowledge N. Cottet and Y. Lu for useful input regarding the manuscript; B. Chapman for providing the parametric amplifiers used for ancilla readout; N. Ofek and P. Reinhold for providing the logic and control interface for the field programmable gate array controller; and C. S. Wang, M.J. Hatridge, and W. Pfaff for valuable discussions.
This research was supported by the U.S. Army Research Office under grants W911NF-18-1-0212 and W911NF-16-1-0349.
L.J. was supported by the Packard Foundation (2013-39273).
Fabrication facilities use was supported by the Yale Institute for Nanoscience and Quantum Engineering (YINQE) and the Yale SEAS clean room.
R.J.S., M.H.D., and L.F. are founders of, and R.J.S. and L.F. are equity shareholders of Quantum Circuits, Inc.

\setcounter{equation}{0}
\setcounter{figure}{0}
\setcounter{table}{0}

\renewcommand{\theequation}{S\arabic{equation}}
\renewcommand{\thefigure}{S\arabic{figure}}
\renewcommand{\thetable}{S\arabic{table}}

\clearpage
\onecolumngrid
\section*{Supplementary Information for: \mytitle}

\section{Experimental setup}\label{sec:setup}

\subsection{Modules}\label{ssec:modules}
\begin{figure*}[t]
    \centering{
    \includegraphics[height=2.5in]{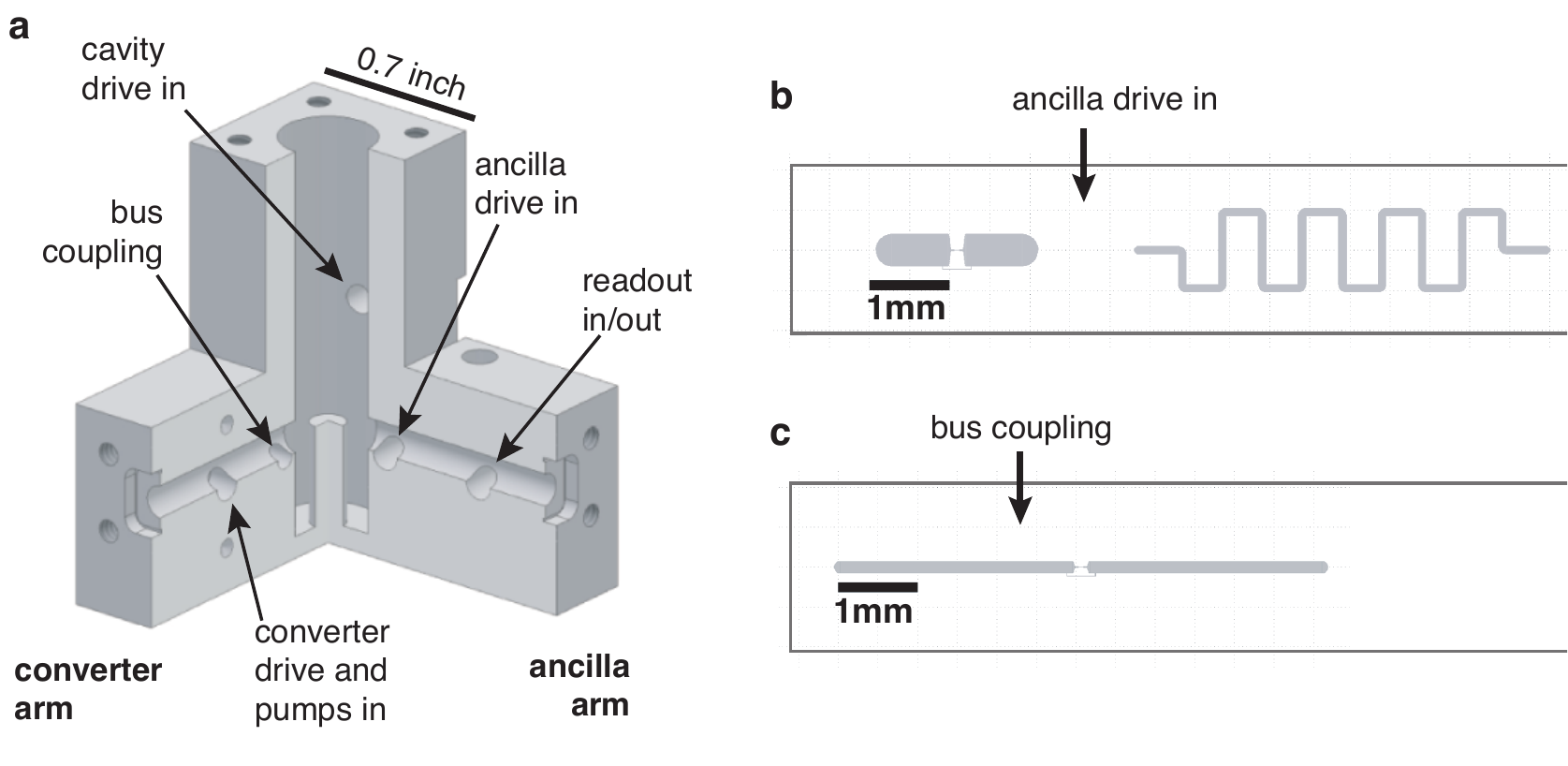}
    \caption[Module and Chip Design]{\label{sfig:sample}
    \textbf{Module and Chip Design.}
    a) Three-quarter cross section of module.
    Post cavity (center) intersected by two tunnels.
    Labels indicate coupling pin locations.
    b) Ancilla chip layout shows transmon pads (left) and meander stripline readout (right).
    Label indicates approximate location of ancilla drive port.
    Readout port and stripline Purcell filter, located further to the right, not shown.
    b) Conversion transmon chip layout.
    Label indicates approximate location of bus coupling pin.
    Drive port, located further to the right, not shown.
    }
    }
\end{figure*}

Each module is a nominally identical device constructed from a solid piece of 99.99\% pure aluminum, chemically etched as in \refref\cite{Reagor2016} to improve surface quality.
Modules consist of a central post cavity \cite{Reagor2016,Axline2016} with two orthogonal tunnels intersecting the cavity, similar to the sample in \refref\cite{Blumoff2016}.
One tunnel houses a chip containing the ancilla transmon, readout resonator, and Purcell filter \cite{Axline2016}.
The other tunnel houses a separate chip with conversion transmon.
This tunnel is intersected by another, smaller tunnel, housing the end of the coaxial bus resonator (see \secref\ref{ssec:bus}.
All chips are double-polished sapphire with aluminum films defined by a single electron-beam lithography step, and double-angle evaporation to form the Josephson junctions in a Dolan bridge process.
Samples are thermally anchored to the base stage of a dilution refrigerator at approximately \SI{20}{\milli\K}, with a magnetic shield surrounding the modules and bus.
Device Hamiltonian parameters and coherence times are listed in \tabref\ref{tab:devparams}.
Module and chip design is shown in \figref\ref{sfig:sample}

\subsection{Coaxial bus resonator}\label{ssec:bus}
\begin{figure*}[t]
    \includegraphics[width = 5in]{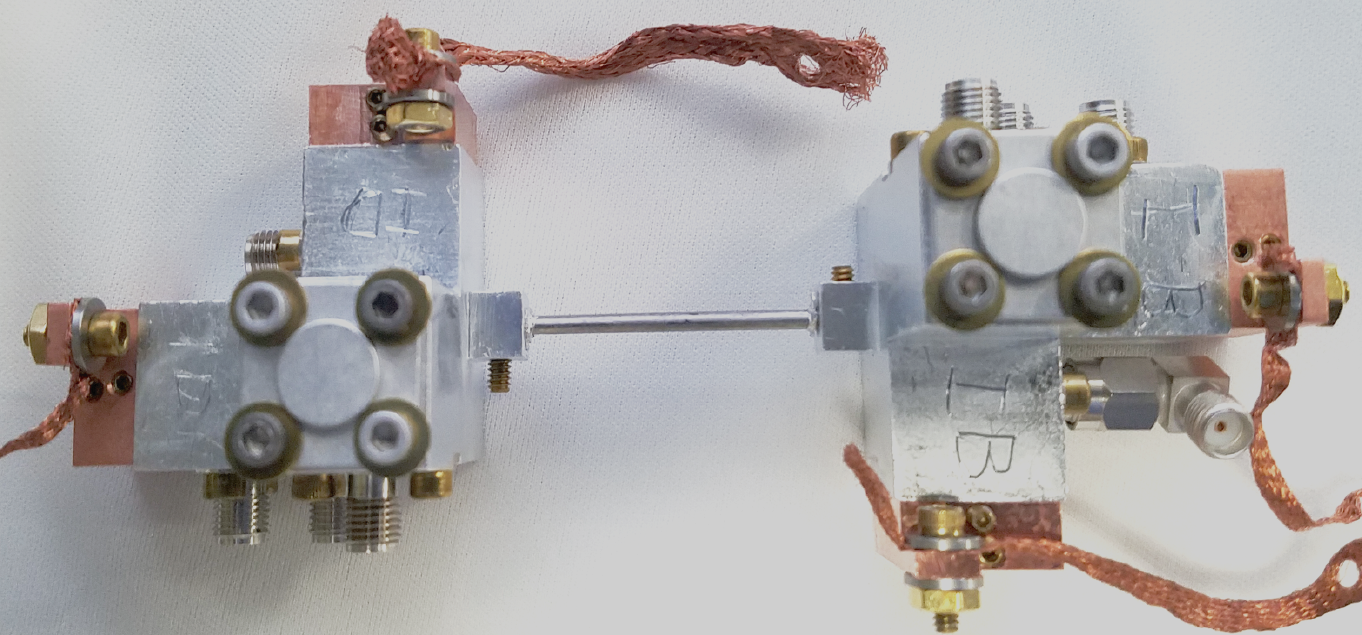}
    \caption{\label{sfig:photo}
    \textbf{Photo of assembled modules.}
    Wiring excluded.
    Sample mounted to copper bracket at base of dilution refrigerator.
    Copper braids connect the copper chip holders directly to the bracket for chip thermalization.
    }
\end{figure*}

Modules are connected by a $l=\SI{6.6}{\cm}$ section of commercial NbTi coaxial cable (Coax Co. SC-086/50-NbTi-NbTi) with PTFE dielectric.
The cable is a multi-mode resonator with free spectral range \SI{1.9}{\GHz}.
The third harmonic ($l = 3\lambda/2$) is used as the bus in this experiment, as it has largest dispersive coupling to the conversion transmon by virtue of being close in frequency.
The final \SI{7.5}{\mm} of outer conductor and dielectric is removed from each end of the cable to expose the inner conductor.
The cable is inserted in a tunnel in the module, with exposed inner conductor near one capacitor pad of the conversion transmon.
The outer conductor is clamped against the body of the aluminum module with a brass screw, with a small amount of bulk indium between screw and cable to provide a larger area of contact force.
This mounting scheme places only two superconducting joints in the path of current flow for the bus mode, one at each end.
The outer conductor is lightly sanded before assembly to remove some of the oxide layer.
The quality of this interface is irreproducible, and likely limits the quality factor of the bus (see \secref\ref{ssec:bus_q}).
\figref\ref{sfig:photo} illustrates the assembled modules and coaxial cable.

\subsection{Experimental wiring}\label{ssec:wiring}
\begin{figure*}[tp]
    \includegraphics{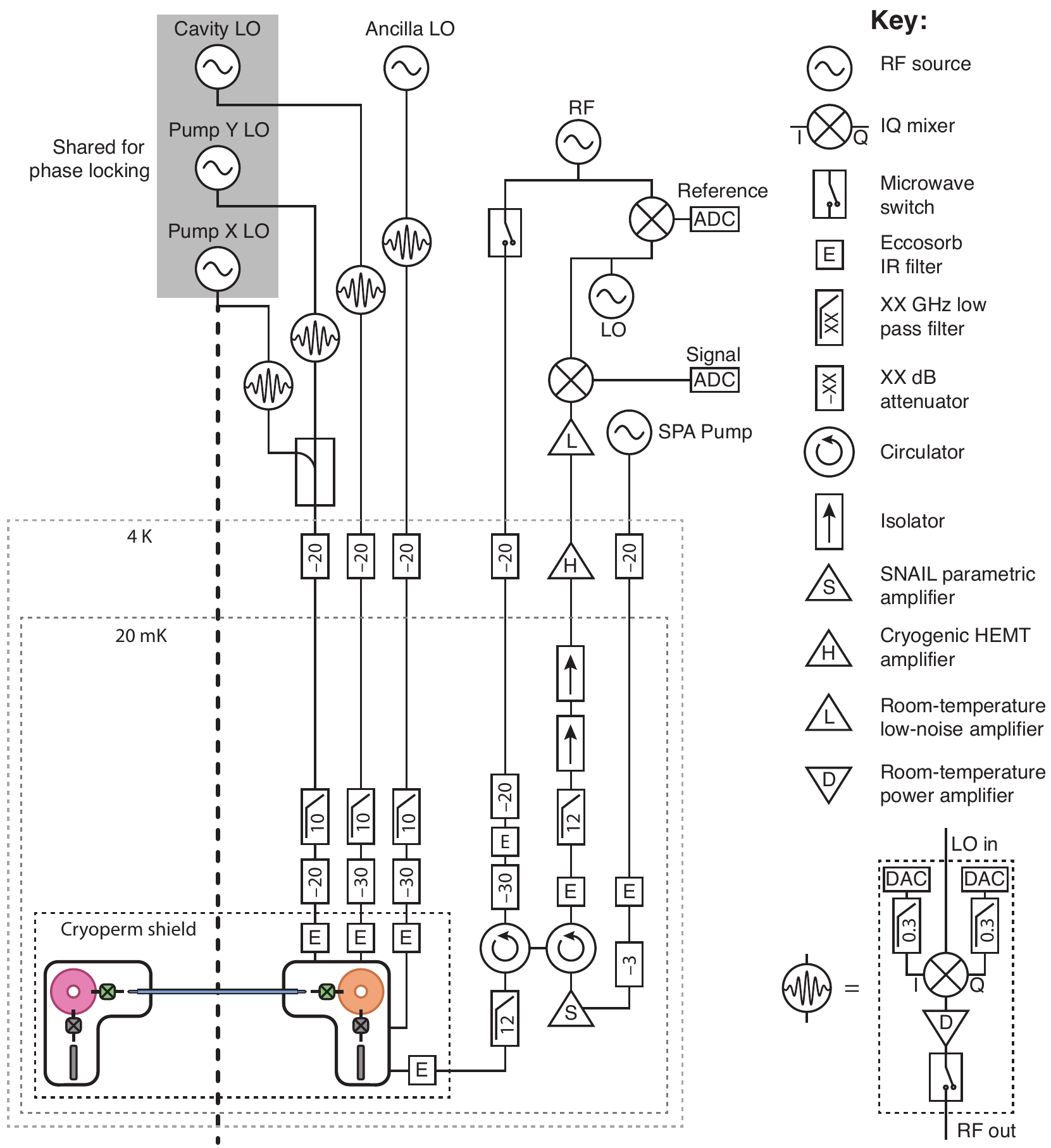}
    \caption{\label{sfig:wiring}
    \textbf{Experimental Wiring Diagram.}
    Input and output signal lines.
    Setup is duplicated across vertical dashed line.
    Three LOs in gray are shared between the two halves for phase-locking.
    Some attenuation and filtering at room temperature not shown.
    }
\end{figure*}

Each module has a mostly separate and identical drive chain to address all modes and apply the off-resonant pumps to the conversion transmon (see \figref\ref{sfig:wiring}).
Measurement of the ancilla transmon is performed with dispersive readout in reflection off a port which is strongly coupled to the Purcell filter.
Readout pulses are sourced at room temperature with a continuous-wave generator (RF) and fast microwave switch.
Reflected readout signals are amplified at the base stage by two SNAIL parametric amplifiers (SPA) \cite{Frattini2018} with approximately \SI{23}{\dB} of gain, \SI{20}{\MHz} of instantaneous bandwidth, and noise visibility ratio of \SI{7}{\dB}.
The SPAs are operated in phase-preserving mode by detuning the pump from twice the readout frequency by about \SI{20}{\MHz}.
SPA pumps are gated by directly pulsing the source generators.
Signals are further amplified at 4 K (\SI{40}{\dB}) and room-temperature (\SI{30}{\dB}), then down-converted to an intermediate frequency (50 MHz) by a separate local oscillator (LO).
This signal, as well as a reference signal formed by mixing the RF and LO continuously, are amplified again (\SI{14}{\dB}) and digitized by a pair of analog-to-digital converters (ADCs).
The signal and reference are compared on each experimental shot, and the relative trajectory is integrated with an appropriate envelope and thresholded to discriminate ancilla states.

All other input signals are IQ modulated by 8 pairs of digital-to-analog converters (DACs), amplified and filtered at room temperature.
DACs, ADCs, and digital channels are on four Innovative Integration X6-1000M cards, which have FPGAs loaded with custom logic.
All control and measurement lines are additionally filtered at low temperature (see \figref\ref{sfig:wiring}).

\subsection{Pump scheme and phase locking}\label{ssec:pump_scheme}
Each conversion transmon is driven by a pair of far-detuned pumps, applied approximately 100 and 1200 MHz above the transmon frequency (called pump X and pump Y, respectively).
Mode and pump frequencies are listed in \tabref\ref{tab:devparams}.
All pump pulses have a constant duration of the length quoted in the text, plus a cosine-shaped rise and fall, each 48 \si{ns} in length.
Since the overall phase of a transmitted state depends on the initial phase as well as the phase of each of the four pumps, three local oscillators are shared between the modules as indicated in \figref\ref{sfig:wiring}.
These local oscillators source the cavity drive, pump X, and pump Y for both modules.
Ancilla and readout input signals are created by independent sources.
Resonant drives for conversion transmons, used only for characterization experiments, are sourced by the same drive chain as pump X.

The use of shared LOs between control setups ensure the LO phases cancel out in the data.
Likewise, the phase of the oscillators which control the single-sideband (SSB) modulation frequency must also be locked.
This is ensured by choosing SSB frequencies such that the parametric conversion frequency condition $\omega_\mathrm{a1} + \omega_\mathrm{X1} - \omega_\mathrm{Y1} = \omega_\mathrm{a2} + \omega_\mathrm{X2} - \omega_\mathrm{Y2}$ is met.
This condition guarantees that a photon converted from cavity 1 into the bus and out into cavity 2 acquires the same phase on every experimental shot.
Additionally, the phase of the oscillators which set these SSB frequencies are reset at the beginning of every experimental shot to remove long-term drifts.

\subsection{Crosstalk between modules}\label{ssec:crosstalk}
While the modules used in this experiment are nominally identical for ease of manufacture, future realizations might benefit from intentional asymmetry between the modules.
The near-degeneracy of the two cavity modes, detuned by only \SI{8}{\MHz}, causes a small amount of crosstalk between control pulses.
We find when we displace one cavity with a wide-band pulse, the other is displaced by 2\textendash{}3\% in amplitude.
This may cause small errors in simultaneous control and tomography, which can be easily mitigated by a small intentional detuning between the two by machining the lengths of the posts to be slightly different.

The strong coupling of the conversion transmons to the bus, and their proximity in frequency space, leads to a strong dispersive coupling to one another.
Due to the situation of the frequencies in the straddling regime \cite{Dicarlo2009}, the dispersive shift between the two conversion transmons is $+2\pi \times \SI{2}{\MHz}$.
Since we do not use the conversion transmons as anything other than a mixing element, this shift plays no role, but it could be used to rapidly entangle the two modules.

\section{Deriving the effective beamsplitter}\label{sec:beamsplitter}

\subsection{Conversion Hamiltonian}\label{ssec:conversion}
The parametric conversion between cavity and bus results from a four-wave mixing process as in \refref\cite{Pfaff2017}.
In the frame of the drives, the effective Hamiltonian describing the interaction between the bus and both cavities can be rewritten as
\begin{equation}\label{seq:bus_ham}
\begin{aligned}
\hat{H}/\hbar &{}={} ig \modea{1} \modeb^\dagger - ig \modea{1}^\dagger \modeb \\
        &{}-{} ig \modea{2} \modeb^\dagger + ig \modea{2}^\dagger \modeb \\
        &{}-{} \Delta \modeb^\dagger \modeb,
\end{aligned}
\end{equation}
We take $g$ to be real and equal for both modules, but the amplitude and phase is controlled by the amplitudes and relative phases of the off-resonant pumps.
In particular, we must tune the amplitudes so that $g$ is equal on both sides.
The common detuning $\Delta$ of the bus mode is controlled by detuning one of the pumps used for each conversion interaction away from the conversion resonance by $\Delta$.
This is the detuning in \figref 2 of the main text.

\subsection{Equations of motion and elimination of the bus}\label{ssec:eom}
The lossless dynamics of this bilinear Hamiltonian may be readily found by solving the Heisenberg equations of motion for $\modea{1}$, $\modea{2}$, and $\modeb$:
\begin{equation}\label{seq:bus_eom}
\begin{aligned}
\dot{\hat{a}}_1(t) &=-g \modeb (t) \\
\dot{\hat{a}}_2(t) &=g \modeb (t) \\
\dot{\hat{b}}(t) &= g (\modea{1}(t) - \modea{1}(t)) + i \Delta \modeb(t)
\end{aligned}
\end{equation}
Since the equations of motion are linear, the solution for the field operators can be written in matrix form as
\begin{equation}\label{seq:bus_sol_matrix}
\begin{pmatrix}
\modea{1}(t) \\ \modea{2}(t) \\ \modeb (t)
\end{pmatrix}
=
\begin{pmatrix}
M_{11} & M_{12} & M_{13} \\
M_{21} & M_{22} & M_{23} \\
M_{31} & M_{32} & M_{33}
\end{pmatrix}
\begin{pmatrix}
\modea{1}(0) \\ \modea{2}(0) \\ \modeb (0)
\end{pmatrix},
\end{equation}
with matrix elements
\begin{equation}\label{seq:bus_sol_matrix_elements}
\begin{gathered}
M_{11} = M_{22}  = \frac{1}{2} \left[ 1 + e^{i\Delta t/2} \left(  \cos(\sqrt{2} \Omega t) - \frac{i\Delta}{\sqrt{8}\Omega} \sin(\sqrt{2} \Omega t) \right)  \right] \\
M_{12} = M_{21}  = \frac{1}{2} \left[ 1 - e^{i\Delta t/2} \left(  \cos(\sqrt{2} \Omega t) - \frac{i\Delta}{\sqrt{8}\Omega} \sin(\sqrt{2} \Omega t) \right)  \right] \\
M_{13} = -M_{31} = -M_{23} = M_{32} = -\frac{g}{\sqrt{2}\Omega} e^{i\Delta t/2} \sin(\sqrt{2} \Omega t)   \\
M_{33} = e^{i\Delta t/2} \left(  \cos(\sqrt{2} \Omega t) + \frac{i\Delta}{\sqrt{8}\Omega} \sin(\sqrt{2} \Omega t) \right) \\
\text{with } \Omega \equiv g\sqrt{1 + \frac{\Delta^2}{8g^2}},
\end{gathered}
\end{equation}
where we have defined the effective interaction rate $\Omega$ in analogy with detuned vacuum Rabi oscillations.
Population thus oscillates between the cavity and bus modes with frequency $2\sqrt{2} \Omega$, with the amplitude of oscillation proportional to $\left(g/\Omega\right)^2$.
In the large detuning limit $\Delta \gg g$, this amplitude  is suppressed, scaling as $\left(g/\Delta\right)^2$, a familiar result in the context of virtual Raman transitions.

This exact solution makes it clear that at time $t=\tau$ such that $\sqrt{2}\Omega \tau = k \pi$ (for integer $k$),  $M_{13} = M_{31} = M_{23} = M_{32}=0$.
This means that $\modea{1}(\tau)$ and $\modea{2}(\tau)$ are decoupled from $\modeb(0)$, and $\modeb(\tau) = \modeb(0)$, up to an overall phase.
This is precisely what we mean by the bus being eliminated.
The first time this elimination occurs ($k=1$) is
\begin{equation}
t = \tbs = \frac{\pi}{\sqrt{2} \Omega} = \frac{2\pi }{\sqrt{8 g^2 + \Delta^2}}
\end{equation}
which we refer to as the beamsplitter time.
The solution for the field operators at the beamsplitter time is
\begin{equation}\label{seq:bs_evolution}
\begin{aligned}
\modea{1}(\tbs) &= e^{-i\theta} \left( \modea{1}(0) \cos \theta + i \modea{2}(0) \sin \theta \right)\\
\modea{2}(\tbs) &= e^{-i\theta} \left( \modea{2}(0) \cos \theta + i \modea{1}(0) \sin \theta \right),
\end{aligned}
\end{equation}
which is a beamsplitter transformation with mixing angle
\begin{equation}\label{seq:bs_angle}
\theta = \frac{\pi}{2} \left( 1 - \frac{\Delta}{\sqrt{8}\Omega} \right) = \frac{\pi}{2} \left(1 - \frac{\Delta}{\sqrt{8g^2 + \Delta^2 }} \right) = \frac{\pi}{2} \left(1 - \frac{\Delta \, \tbs}{2\pi} \right).
\end{equation}
This general result establishes the two working cases used in the main text, explained on the next two subsections.

\subsection{State transfer and efficiency}\label{ssec:transfer}
The first useful working point is used for state transfer.
By choosing $\Delta = 0$, we have $\theta = \pi/2$, which occurs at time $\tswap = \pi / (\sqrt{2} g)$.
This results in evolution
\begin{equation}\label{seq:swap_evolution}
\begin{aligned}
\modea{1}(\tswap) &=  \modea{2}(0) \\
\modea{2}(\tswap) &=  \modea{1}(0),
\end{aligned}
\end{equation}
which swaps the state of the two cavities.
Our choice of the phase of $g$ in \eqnref{seq:bus_ham} results in a true SWAP operation at these conditions.
Any deviation in the phase of $g$ would result in a cavity phase space rotation on one or both of the input states, which can be calibrated out in any encoding \md this is not a logical qubit phase.

In this case, we can easily include the effect of damping in the bus by replacing $\Delta\rightarrow i\kappab/2$ in \eqn\ref{seq:bus_eom}.
In this case, we replace the detuned Rabi frequency $\Omega$ with the loaded oscillation frequency $\tilde{g} = g\sqrt{1 - \kappab^2/(32g^2)}$.
To compute the energy efficiency in the presence of loss, it is sufficient to consider the case of an input coherent state $\ket{\alpha}$ in cavity 1, and vacuum in cavity 2, which emulates the one-way state transfer demonstrated in the main text.
In this semi-classical case, it is convenient to replace the field operators with their expectation values, e.g. $a_j(t) = \ev{\modea{j}(t)}$.
The initial conditions are then $a_1(0)=\alpha$ and $a_2(0)=b(0)=0$.
As long as the oscillations are under-damped ($\kappab < \sqrt{32}g, \tilde{g} \in \mathbb{R}$), the bus will still periodically be eliminated.
The exact solution for the dynamics is given by
\begin{equation}\label{seq:bus_resonant_loss}
\begin{aligned}
a_1(t) = \ev{\modea{1}(t)} &= \frac{\alpha}{2} \left[ 1 + e^{-\kappab t/4} \left(  \cos(\sqrt{2} \tilde{g} t) + \frac{\kappab}{\sqrt{32}\tilde{g}} \sin(\sqrt{2} \tilde{g} t) \right)  \right] \\
a_2(t) = \ev{\modea{2}(t)}&= \frac{\alpha}{2} \left[ 1 - e^{-\kappab t/4} \left(  \cos(\sqrt{2} \tilde{g} t) + \frac{\kappab}{\sqrt{32}\tilde{g}} \sin(\sqrt{2} \tilde{g} t) \right)  \right] \\
b(t) = \ev*{\modeb(t)}   &= \frac{\alpha}{\sqrt{2}} \frac{g}{\tilde{g}} e^{-\kappab t/4} \sin(\sqrt{2} \tilde{g} t) .
\end{aligned}
\end{equation}
For the values used in this experiment, the loading of the oscillation frequency is a very small effect, and $\tilde{g}\approx g$ to a very good approximation.

The energy efficiency of the transfer, $\eta$, is the energy in mode $a_2$ at the end of the transfer relative to the initial energy in mode $a_1$, and is given by
\begin{equation}\label{seq:efficiency}
\eta = \frac{|a_2(\tswap)|^2}{|a_1(0)|^2} \approx \frac{1}{4} \left(1+e^{\frac{-\pi\kappab}{\sqrt{32} g}}\right)^2 \approx 1 - \frac{\pi \kappab}{\sqrt{32} g}
\end{equation}
which is $\eta = 0.89$ for the experimentally measured values of $g$ and $\kappab$ quoted in the main text.

\subsection{Entanglement generation}\label{ssec:entanglement}
The other regime utilized in this work is the 50:50 beamsplitter ($\theta=\pi/4$), which is obtained at $\Delta = g\sqrt{8/3}$.
At this detuning, $\Omega = g\sqrt{4/3}$, and the interaction time is $\tfty=\pi\sqrt{3/8}/g = \sqrt{3/4}\tswap$.
The resulting operation is
\begin{equation}\label{seq:bs_evolution_fty}
\begin{aligned}
\modea{1}(\tbs) &= \frac{1}{\sqrt{2}}e^{-i\pi/4} \left( \modea{1}(0) + i \modea{2}(0) \right)\\
\modea{2}(\tbs) &= \frac{1}{\sqrt{2}}e^{-i\pi/4} \left( \modea{2}(0) + i \modea{1}(0)  \right),
\end{aligned}
\end{equation}
which can produce maximally entangled final states for certain initial states.

Loss in the bus will also introduce a finite efficiency in the $\theta = 45^\circ$ beamsplitter.
For a singe-photon input state, this efficiency is to leading order in $\kappab/g$
\begin{equation}\label{seq:inefficiency_ent}
1 - \eta_\mathrm{BS} =\frac{\kappab g^2}{2\Omega^2}\frac{\tfty}{2} =  \frac{3\sqrt{3}\pi}{8\sqrt{32}} \frac{\kappab }{g} \approx 0.36 \frac{\kappab}{g},
\end{equation}
which is less loss than the resonant swap because the process is faster and populates the bus less.

Since there is only one excitation, the state which results from loss is $\ket{00}$.
So the above inefficiency results in a mixture of the ideal Bell state and the vacuum state:
\begin{equation}
\rho = \eta_\mathrm{BS} \frac{1}{2}\left( \ket{01} + \ket{10} \right)\left( \bra{01} + \bra{10} \right) + (1-\eta_\mathrm{BS}) \ketbra{00}
\end{equation}
The second term has zero fidelity to the ideal state, so
\begin{equation}
1 - F_\mathrm{Bell,bus} = 1 - \eta_\mathrm{BS} = \frac{3\sqrt{3}\pi}{8\sqrt{32}} \frac{\kappab }{g}
\approx 0.36 \frac{\kappab }{g}.
\end{equation}

\subsection{Alternate entangling schemes}
There are other many ways the conversion process used here can generate entanglement.
For instance, modulating the conversion couplings in time can effect an entangling partial swap.
One such approach is to prepare a single photon in cavity 1, turn on conversion to the bus for only cavity 1 for time $t_{\text{half}} = \pi /(4g)$ to ``half-swap'' a single photon into the bus.
This creates a Bell pair between cavity 1 and the bus.
We may then turn on conversion from the bus to cavity 2 for $t_{\text{full}} = \pi/(2g)$, which fully swaps the bus occupation into cavity 2, resulting in a Bell state between the two cavities.
A continuous version of this protocol involves simultaneous on resonance conversion with unequal strength, such that $g_1 = \left(\sqrt{2}-1\right)g_2$.
Dynamics are qualitatively similar to the beamsplitter operation we use in this work, with the bus mode being occupied at intermediate times and returning to vacuum after time ${\sim}\pi/g$ resulting in the 50:50 beamsplitter relations.
However, such protocols implement Hamiltonians that are not symmetric under the exchange of modes $a_1$ and $a_2$.
As a consequence, should a photon loss event occur in the bus mode, the environment will gain information that projects the cavities into states that are not symmetric in $a_1$ and $a_2$, making it difficult to implement robust multi-photon entanglement schemes.

For protocols such as the error-detected Hong\textendash{}Ou\textendash{}Mandel entanglement scheme, we must make sure to engineer a ``true beamsplitter'' transformation.
For this scheme, if photon loss occurs in the bus mode, the environment does not learn from which cavity this excitation originated, due to the indistinguishability implied by the symmetry of the interaction.
In fact, even after a single photon loss event in the bus, the joint cavity state is ideally the single-photon Bell state $\ket{01}+\ket{10}$.
It is only after the parity measurements that the state is projected into either $\ket{01}$ or $\ket{10}$.
Entanglement schemes that are insensitive to photon loss on the bus or otherwise rely on it will be the subject of future work.
Another approach robust state transfer and entanglement generation is stimulated Raman by adiabatic passage \cite{Bergmann1998}, which modulates the coupling strengths in time in a way which suppresses the occupation of the bus at all times.
However, since this modulation must be adiabatic with respect to the maximum coupling strength, such protocols are necessarily much slower than the ones used here, similar to the virtual Raman approach.
Loss induced by non-adiabaticity will not have the indistinguishability properties of the beamsplitter.

Additionally, encoding-independent entangling gates between bosonic modes such as exponential-SWAP and Fredkin gates \cite{Gao2019} may be constructed by sandwiching local operations with ancillae between 50:50 beamsplitter operations.
The tools demonstrated in this work enable such operations between separable modules.

\section{Bus resonator characterization}\label{sec:bus_char}

\subsection{Quality factor and attenuation length}\label{ssec:atten_and_q}
With quality factors in the tens of thousands, the hypothetical maximum state transfer efficiency from just using a section of superconducting coaxial cable is extremely high.
To illustrate this point, we consider the \SI{6.6}{\cm} section of cable we use for this work, which has modes with $Q\approx \num{50000}$.
Regarding the cable as a Fabry-Perot cavity with uniform loss, this quality factor corresponds to an energy attenuation length of $\SI{300}{\m}$ \cite{Kurpiers2017a}.
Over reasonable meter-scale lengths of cable within a cryostat, this corresponds to a single-pass loss approaching $10^{-3}$.
This fundamental limit is orders of magnitude smaller than the single pass loss observed in circulator-based communication links, which are limited to around $0.1$ \cite{Axline2018,Campagne-Ibarcq2018,Kurpiers2018}.

In the current implementation, the achievable efficiency is limited by the speed of the protocol (see \eqnref{seq:efficiency}), as a transmitted photon essentially makes many passes through the bus.
However, by increasing the conversion strength would move closer towards this fundamental limit.
Furthermore, as the bus is a 3D cavity, it is reasonable to believe that the quality factor can be improved by several orders of magnitude with improved materials and better seam quality.

\subsection{Quality factor measurements}\label{ssec:bus_q}

\subsubsection{Measuring the bus \emph{ex situ}}
\begin{figure*}[t]
    \centering{
    \includegraphics{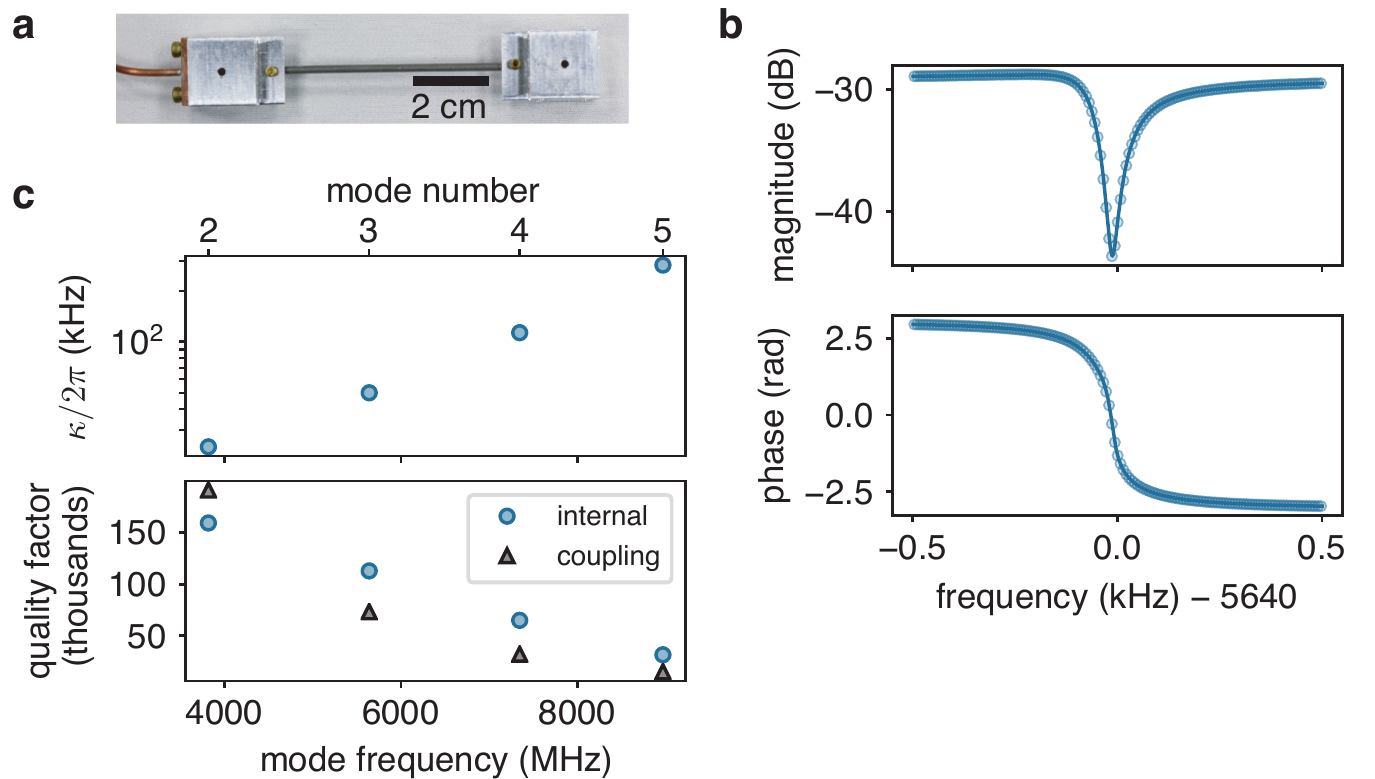}
    \caption[Bus Characterization \emph{ex situ}]{\label{sfig:cable_vna}
    \textbf{Bus Characterization \emph{ex situ}.}
    a) Setup for measuring cable resonators in reflection.
    Cable is embedded in aluminum tunnel, held in place with brass set screw.
    Copper cable (left) enters other end of tunnel, and is soldered into a flange.
    Both have outer conductor and dielectric removed in the tunnel (not shown).
    b) Reflection measurement of $n=3$ mode, with fit.
    c) Extracted internal decay rate $\kappa$ and internal and coupling quality factors for four modes of the same cable.
    }
    }
\end{figure*}

We have used a scheme similar to \refref\cite{Kurpiers2017a} to characterize the quality of the bus mode before integrating it with the modules.
We couple to the modes of a section of cable in reflection and measure $S_{21}$ with a vector network analyzer.
The cable is terminated at one end in a tunnel of 6061 aluminum alloy that functions as a waveguide below cutoff.
At the other end, we capacitively couple to the cable via a coupling pin in a similar tunnel.
We set the distance between the coupling pin and center conductor of the cable such that we are nearly critically-coupled.
The cable is secured to these blocks of aluminum in the same way it is attached to the modules, with an indium-tipped brass set screw.
Typical quality factors are of order $\sim 50,000$ although quality factors as high as $\sim 160,000$ have been observed.
Quality factors can differ by factors of 2 to 3 in the same cable when we re-set the mounting screws between cooldowns.
As such, we suspect seam loss between the outer conductor of the cable and the aluminum package of the modules to be a limiting factor.
Efforts are ongoing to improve the quality and reproducibility of the seam between the cable and modules.
This screening process also allows us to measure the frequency and quality of potential bus resonators before assembling the full experimental hardware.
We are also able to measure multiple modes of the same cable in this way.
\figref\ref{sfig:cable_vna} shows such measurement of four modes of the same cable used in the main text.

\subsubsection{Measuring the bus \emph{in situ}}
\begin{figure*}[t]
    \centering{
    \includegraphics{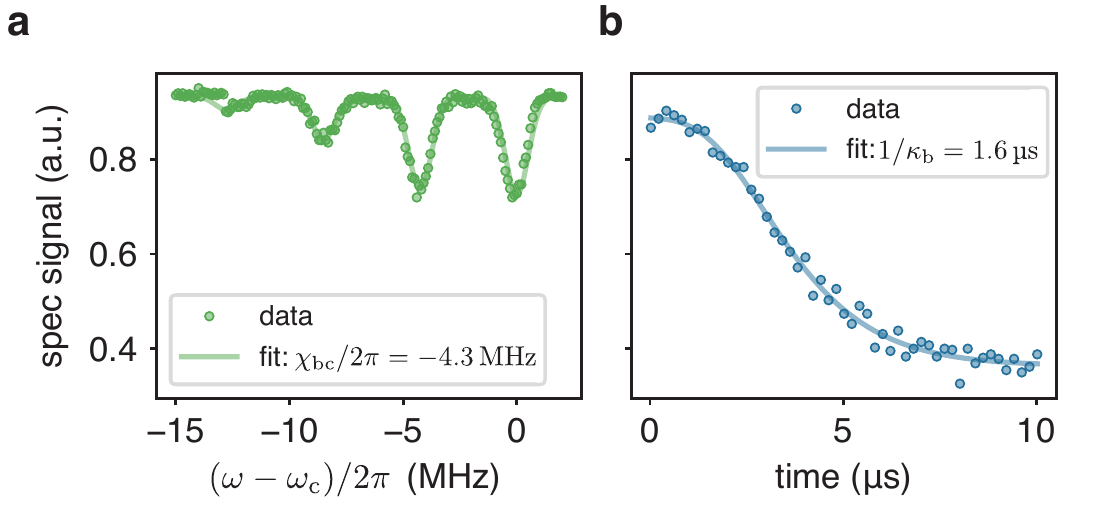}
    \caption[Bus Characterization \emph{in situ}]{\label{sfig:cable_in_situ}
    \textbf{Bus Characterization \emph{in situ}.}
    a) Spectroscopy on conversion transmon for a coherent state $\ket{\alpha=1}$ in the bus.
    b) Lifetime measurement of bus.
    Bus is displaced to $\ket{\alpha=2.5}$, then after a variable delay, the conversion transmon is driven with a selective $\pi$ pulse at $\omega = \omega_\mathrm{c}$.
    }
    }
\end{figure*}

When the bus is installed in the modules, we can measure its properties without any dedicated drive or measurement lines.
Since the mode of interest has a dispersive shift to the conversion transmon, we can detect population in the bus by performing spectroscopy on the converter, much the way we measured storage mode population with the ancilla.
The results of this spectroscopy with the cable driven to a coherent state is shown in \figref\ref{sfig:cable_in_situ}a.

Given the number-resolved spectrum in \figref\ref{sfig:cable_in_situ}a, the damping rate of the cable $\kappab \ll \chibc$.
We can measure this rate directly by displacing the cable and driving the converter with a selective pulse at $\omega = \omega_\mathrm{c}$.
The height of this spectroscopic peak corresponds to the occupation of the $\ket{0}$ state of the bus.
This ring-down measurement, shown in \figref\ref{sfig:cable_in_situ}b, reveals a bus lifetime of \SI{1.6}{\us}, or $\kappab/2\pi = \SI{100}{\kHz}$.

\begin{figure*}[t]
    \centering{
    \includegraphics{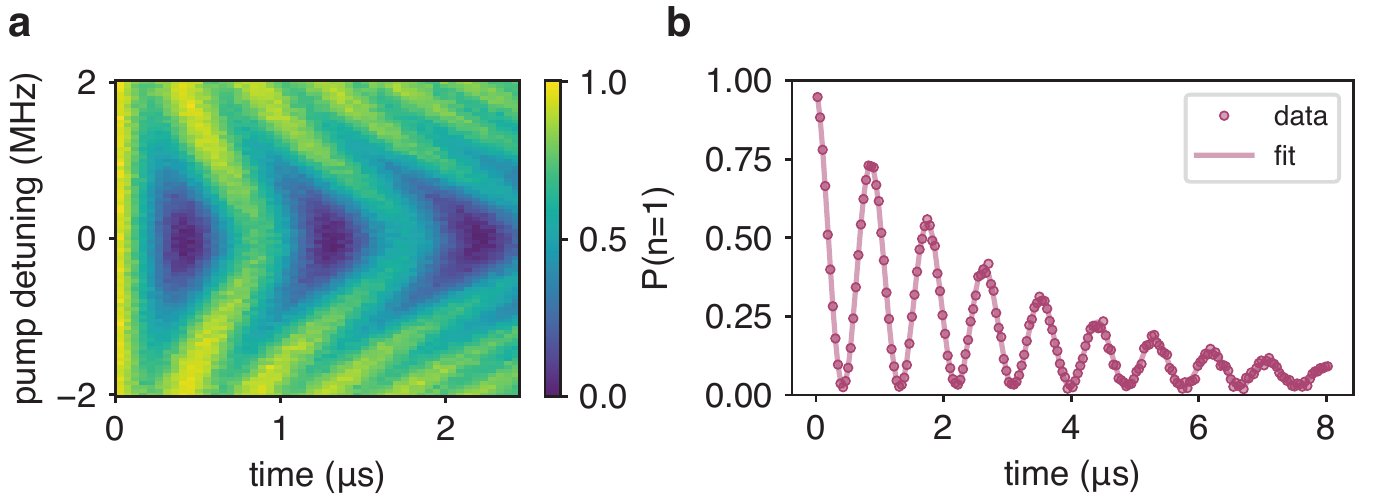}
    \caption[Cavity-Bus Swapping Dynamics]{\label{sfig:chevron}
    \textbf{Cavity-Bus Swapping Dynamics.}
    a) A single photon is prepared in cavity 1, then conversion to the bus is enable by turning on both pumps.
    Measured occupation of $\ket{1}$ in cavity 1 shown versus time and detuning of one pump from resonance.
    b) Line cut at zero detuning shows ring-down of population from decay in the bus.
    Value of conversion strength and decay rate in text is extracted from fit.
    }
    }
\end{figure*}

We certify that the lifetime of the bus is not degraded in the presence of the conversion pumps.
This is done by preparing a single excitation in one of the storage modes and turning on conversion on only one module, such that excitations swap into the bus and back into the storage cavity.
By measuring the storage population after applying conversion for variable time $t$ we can extract both $g$ and $\kappa$ from the fit in \figref\ref{sfig:chevron}.
The fitted value $\kappab/2\pi = \SI{110}{\kHz}$ shows minimal change in the presence of the conversion process.
Likewise, fitting the data in \figref 2b of the main text results in a similar value of $\kappab$.
We also verify that the cable has no measurable thermal population by examining the storage population after a single swap between the storage and bus when we initialize vacuum.

\subsection{Dependence of scheme on cable length}
The section of cable we use is fairly short, suitable for joining adjacent modules together.
For certain applications, it may be beneficial to have meter-scale lengths of cable which can connect any two modules within a cryostat.
For all other hardware kept fixed, there are some challenges which arise when making the link much longer.
Firstly, the coupling rate $g$ would decrease due to a reduction in the cross Kerr between the conversion transmon and the bus mode.
This cross-Kerr is proportional to the energy participation of the bus mode in the junction of this transmon.
From a simple volume argument, the energy density in the bus mode should scale as $\sim 1/l$ and thus so should the cross-Kerr.
The parametric conversion strength $g$ scales as the geometric mean of this cross-Kerr and the cross-Kerr between the storage cavity and the conversion transmon, so we expect $g\sim 1/\sqrt{l}$.

This reduction in $g$ may be compensated by increasing the capacitive coupling between the conversion transmon and the end of the cable.
In doing so, we may Purcell limit the transmon and storage cavity as well as increase crosstalk between modules.
We have reason to believe we may already be Purcell limiting the conversion transmons \textemdash{} their lifetimes as measured before the addition of the bus were a factor of a few longer.

The Purcell effect arises due to off-resonance static couplings to modes of the cable even in the absence of conversion drives.
As we make the cable longer, the free spectral range decreases as $1/l$.
Within a fixed frequency window, there are now more cable modes the conversion transmon or storage mode can couple to off-resonantly.
This potential issue could be addressed with filtering dedicated filter modes between the conversion transmon and the bus.
This is similar to the approached used in \refref\cite{Leung2019}, but in this case the filter modes need not be precisely frequency-matched.
Improving the quality of the bus modes would also lessen the effect of the Purcell limit.

Finally, it is worth emphasizing that even for a longer cable, we can still effectively model the dynamics by considering coupling to a single bus mode and the protocols would be unchanged.
Only when $g \gtrsim \text{FSR}$ will we need to consider simultaneous coupling to multiple bus modes, as was explored in \refref\cite{Zhong2019}.
For a \SI{1}{\m} cable, the FSR is $\sim \SI{100}{\MHz}$, much larger than the conversion strengths we can engineer at present.

\section{Experimental and analytical techniques}

\subsection{Measurement}\label{ssec:msmt}

\subsubsection{Ancilla measurement}\label{sssec:readout}
For all measurements which are not at the end the experimental sequence, we use a \SI{460}{ns} long square readout pulse calibrated to discriminate ancilla states $\ket{g}$ and $\ket{e}$.
The acquisition window is \SI{580}{ns} long.
This measurement is used for system reset as well as parity assignment.
Ancillae have assignment fidelities of $0.99$ for $\ket{g}$ and $0.98$ for $\ket{e}$, with the asymmetry  due to relaxation events during the measurement.

\subsubsection{Tomographic readout}\label{sssec:tomo_readout}
Measurements used for tomography (the final measurements in an experimental run) are preceded by a pulse on the ancilla which inverts the population in $\ket{e}$ and $\ket{f}$.
The readout pulse and acquisition (500 and \SI{640}{ns}, respectively) are longer, and calibrated to distinguish states $\ket{g}$ and $\ket{f}$.
This allows for a measurement which is much less sensitive to decay events of the ancilla \cite{Mallet2009,Elder2020}, providing higher and more symmetric assignment fidelity, higher than $0.995$ for all states.
This improves the measurement contrast and reduces errors in the single-shot projective measurement in the entangled state tomography (\ssecref\ref{ssec:entangled_tomo}).

\subsubsection{Cavity and ancilla manipulation}\label{sssec:manipulation}
Unless otherwise noted, all ancilla rotations are effected with 40 ns Gaussian pulse ($\sigma = 10$ ns).
Cavity displacements are \SI{40}{ns} Gaussian pulses ($\sigma = \SI{10}{ns}$).
All other manipulation of the cavity state are carried out with numerically-optimized control pulses (NOCP) on the cavity and ancilla using the GRAPE algorithm \cite{Heeres2017}.
Pulse lengths are 500-1200 \si{ns} depending on the operation.

\subsubsection{Parity measurement}\label{sssec:parity_msmt}
Cavity photon number parity measurement is effected with a Ramsey-type sequence on the qubit \cite{Sun2014a}.
Two $90^\circ$ ancilla rotations with an inter-pulse delay of $\pi/\chiat=$ 416 (492) \si{ns} for module 1 (2) entangles ancilla state with photon number.
Ancilla rotations for parity measurement are 24 \si{ns} Gaussian pulses ($\sigma = 4$ ns) to make pulses maximally unselective on photon number.
Parity measurements used for error detection have the phase of the second rotation reversed to map even number (the most probable outcome) onto the ground state of the ancilla, to minimize the probability of errors during ancilla measurement.

\subsubsection{Cavity population measurement and normalization}\label{sssec:pop_msmt}
Measurement of the occupation of the $n$th Fock state in the cavity is made by applying a spectrally narrow rotation on the ancilla, at frequency $\omega_q+n\chi$, exciting the ancilla only for this number state.
Selective pulse lengths are \SI{1200}{ns} ($\sigma = \SI{400}{ns}$) for module 1, \SI{1920}{ns} ($\sigma = \SI{480}{ns}$) for module 2.
To normalize for errors in the pulse and readout, we measure this occupation for each relevant $n$, as well as a reference measurement of the ancilla state with no rotation, then subtract the reference and normalize so that the sum of all occupations is 1.
This normalization procedure is applied to the data in Figure 2b,c in the main text.

\subsubsection{Conversion transmon measurement}\label{sssec:blind}
The conversion transmons do not have their own readout resonators, and are measured indirectly through the cavities.
This is used only for system reset and characterization measurements.
The cavity can be used to measure the converter as in \refref\cite{Blumoff2016}.
After the cavity is determined to be in its vacuum state (see Section \ref{sec:reset}), it is displaced to a coherent state with amplitude $\alpha$, typically $\sim1.5$.
After a delay of $\sim 200$ \si{ns}, the opposite displacement is applied.
If the converter is in its ground state, the coherent state will not have moved during the delay, and will return to vacuum.
If the converter was excited during this time, the cavity state will rotate by an angle $\chiac t$, typically $\sim 70^\circ$.
The reverse displacement brings this to another coherent state with a very small overlap with the vacuum state.
By applying a $\pi$-pulse to the ancilla, selective on zero photons in the cavity, we obtain an excitation probability proportional to the converter excitation probability, with a fidelity of about 0.95.
Importantly, this measurement is unlikely to incorrectly give the result corresponding to the converter in its ground state, so it is useful for verifying with high confidence that the converter is not excited.

\subsection{System reset and state preparation}\label{sec:reset}

\begin{figure*}[tp]
    \includegraphics{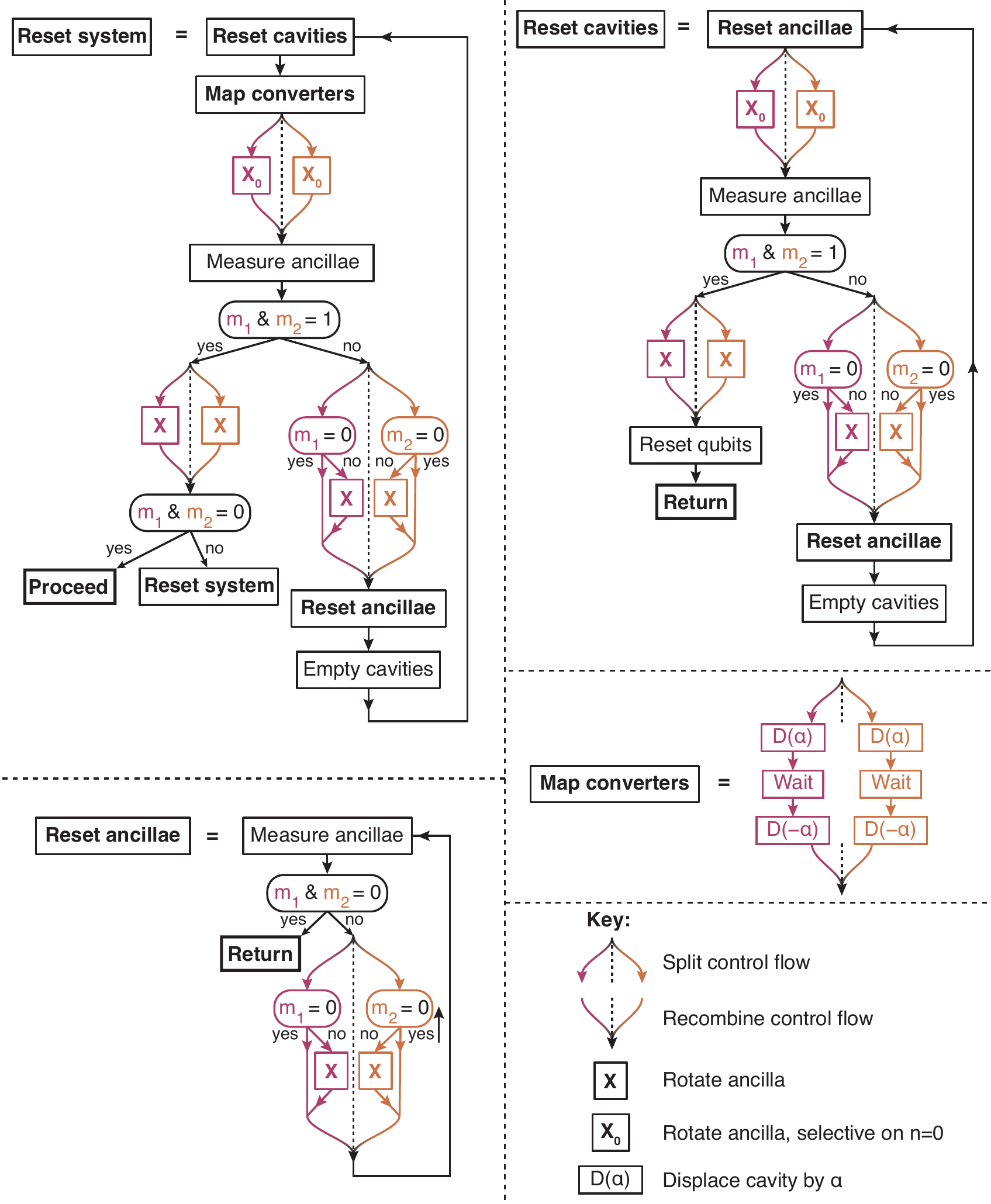}
    \caption{\label{sfig:reset}
    \textbf{System Reset.}
    Logical control flow for system reset and verification.
    Each experimental sequence begins with a call to ``Reset System,'' which follow the control flow until the conditions have been met to reach ``Proceed.''
    Individual subroutines loop until reaching ``Return,'' at which point they return to the previous sequence.
    }
\end{figure*}

\subsubsection{System reset by feedback}
To ensure the modules begins in a known state, we use an active feedback cooling sequence that makes use of the ability of our control hardware to perform simultaneous and independent control branching when resetting the ancilla transmons.

The set of nested subroutines used at the beginning of every experimental sequence is shown in \figref\ref{sfig:reset}.
The sequence begins by ensuring both ancillae are in their ground states, actively resetting as necessary.
Then a $\pi$ pulse, selective on $n=0$ photons in the cavity, is applied to each ancilla, followed by measurement.
Measurement of the ancilla in $\ket{e}$ heralds an empty cavity.
If both cavities are empty, we continue (see next paragraph).
If not, we reset the ancillae to $\ket{g}$, then actively empty the cavities by performing swaps with the bus (as in \figref\ref{sfig:chevron}), one at a time, with a \SI{10}{\us} delay after the swap to allow the state to decay in the bus.
This is at least two orders of magnitude faster than waiting for the long-lived cavities to decay on their own.
We then start the sequence over, beginning with ancilla reset, and repeating as necessary to ensure both cavities are in the vacuum state.

Once the cavities are confirmed empty, we use them to measure the state of the conversion transmons.
This uses the Ramsey-style selective displacement described in \secref\ref{sssec:blind}, which displaces the cavity if the transmon is not in its ground state.
We then repeat the cavity measurement.
If the cavities are again found in vacuum, we know the converters were in $\ket{g}$.
We then reset the ancillae one last time, and begin the experimental sequence.
If either cavity is not in the vacuum state, this means its converter was not in $\ket{g}$.
We then empty both cavities and begin the entire sequence from the beginning.
Since the time to empty the cavities is relatively long, we do not actively reset the converters, but simply allow them to decay during this time.

After successful completion of this cooling routine, we find the ancilla transmons and cavities with less than 1\% probability each of being out of their respective ground states.
The conversion transmons are difficult to measure to this degree of accuracy, since their measurement sequence is fairly long and involved.
Given the length of this sequence, it is likely that they have re-thermalized to about the 1\% level (each) by the time the cooling is successfully completed.

\subsubsection{State preparation}\label{ssec:prep}
As discussed above, all nontrivial cavity states are prepared with NOCPs.
For all operations used to prepare cavity states, the ancilla is meant to return to its ground state at the end of the the pulse.
Errors during the operation can result in occupation of the excited state, usually with probability $2-3\%$ depending on the pulse.
To detect these errors, we measure the state of the ancilla after application of the pulse for all experiments.
If not measured to be in its ground state, we consider this a failure of the preparation, and reset the entire system before trying again.
This measurement is responsible for the small, deterministic phase shift seen in the Wigner tomograms of the states as prepared, shown in \figref 3b,c of the main text.
While this makes the entire experiment probabilistic, we regard this as a part of the system initialization process, which is already nondeterministic.
Operations not at the beginning of the experimental sequence are not error-detected in this way.
With improved ancilla coherence and calibration, this step is not necessary \cite{Heeres2017}.

\subsubsection{Logical state encoding}\label{sssec:encoding}
For each of the two logical encodings used in Figure 3 of the main text, the state is encoded in the cavity using NOCPs.
First the state is prepared in the ancilla transmon with a phase- and ampltiude-controlled rotation.
Then the pulse, which maps combined ancilla-cavity state $\ket{g}\ket{0}$ ($\ket{e}\ket{0}$) onto $\ket{g}\ketL{0}$ ($\ket{g}\ketL{1}$).
As stated in Section \ref{ssec:prep}, we then confirm the ancilla has successfully returned to its ground state before proceeding.
The mean fidelity of the encoded states are $0.99$ for the Fock encoding and $0.98$ for the cat encoding, obtained from Wigner tomography.

\subsection{Wigner tomography and reconstruction}\label{ssec:wigner}

\subsubsection{Measurement, symmetrization, normalization and reconstruction}
The Wigner function measurement is carried out as in \refref\cite{Sun2014a}, for instance.
The cavity is displaced by a variable amount $\beta$, and the average parity is measured using the parity measurement described in \secref\ref{sssec:parity_msmt}.
To symmetrize the measurement, we perform two distinct parity mapping sequences \md one which maps even photon numbers to $\ket{g}$ of the ancilla, and one which maps even to $\ket{e}$.
We take the difference of the two resulting datasets.
This symmetrizes the Wigner function against biased readout errors and finite number-selectivity of the ancilla rotations.

The Wigner function of any physical cavity state should integrate to 1, even for a mixed state.
Since our reconstruction routine assumes the data to be physical, we normalize the measured Wigner functions by a trapezoidal 2D integral over the entire dataset.
This corrects for loss of contrast due to the parity mapping sequence and the ancilla measurement, an effect of 2\textendash{3}\%.
This results in the data presented in the main text.

The cavity state $\rho$ is reconstructed from the Wigner function using is a maximal likelihood estimation, the same routine used in \cite{Axline2018}.
The routine is a convex optimization over the space of physical cavity density matrices with dimension $d=8$ for all data.
Since the largest states measured have mean photon number $\bar{n} \le 2$, this Hilbert space is sufficiently large to capture all population.
The physicality constrains are that $\rho$ is positive semi-definite and $\Tr{\rho}=1$.

\subsubsection{Fidelity error bars}
The state fidelities quoted in the text are computed as the fidelity of the reconstructed state to the ideal state $\rho_\mathrm{ideal}$, $F = \Tr{ \sqrt{  \sqrt{\rho_\mathrm{ideal}} \rho  \sqrt{\rho_\mathrm{ideal}} } }^2$.
The net contribution of errors in reconstruction due to noise and systematic errors such as the dependence of the parity measurement contrast on mean photon number contribute about 1\% error on average, as estimated from simulating these imperfections on ideal data.
This gives the error bar quoted for most of the mean state fidelities in the text.
The systematic error for $\bar{F}_\text{odd}$, which is reconstructed from Wigner functions taken after a measurement of odd parity, is larger due to the occurrence of ancilla decay errors during the first parity measurement, since odd parity is associated with a measurement of the ancilla in $\ket{e}$.
These errors result in distortion of the measured Wigner function from to dephasing caused by the dispersive shift to the ancilla.
However, due to the low probability of this case, the overall error in $\bar{F}_\text{cat,tracked}$ is not as large.

The errors in the entangled state reconstruction, described in \secref\ref{ssec:entangled_tomo}, are similar, since this method relies mostly on density matrices reconstructed from Wigner functions.

\subsection{Entangled state tomography}\label{ssec:entangled_tomo}
To clearly illustrate the correlations between the two cavities, we perform Wigner tomography on cavity 1, post-selected on a logical measurement in cavity 2 in the $x,y,$ and $z$ bases, as indicated in \figref 4a,d of the main text.
Since the Wigner function is a complete description of the state, these conditional Wigner tomograms provides enough information to reconstruct the full two-qubit state.

\subsubsection{Logical basis measurement}\label{sssec:basis_msmt}
The logical basis measurements for entanglement characterization are effected by decoding the cavity state onto the ancilla using NOCPs.
These are the opposite of the encoding operations in \secref\ref{sssec:encoding}.
The mapping is $\ket{g}\ketL{0}$ ($\ket{g}\ketL{1}$) to $\ket{g}\ket{0}$ ($\ket{e}\ket{0}$).
We then measure the ancilla to effect a $z$ basis measurement, or rotate the ancilla into the appropriate basis with a $\pi/2$ pulse around the Y (X) axis to measure in the $x$ ($y$) basis.

To assess the fidelity of the decoding operations, we prepare six cardinal states in the ancilla ($\ket{{\pm}z}$, $\ket{{\pm}x}$, and $\ket{{\pm}y}$), encode and immediately decode, then apply the rotation which should restore the ancilla to the ground state, and measure.
We find on average a 3\textendash{}4\% error, depending on the encoding.
We assume the infidelity of encoding and decoding is similar, and attribute half the average incorrect measurement result to the ``decode, rotate, and measure'' operation, which is the same operation that makes up the logical basis measurement.
This yields the $\sim 2\%$ tomographic error quoted in the main text.

\subsubsection{Two-qubit state reconstruction}
Conditional Wigner tomograms for the $z$ and $x$ bases are shown in \figref 4b,e of the main text; the complete dataset is shown in \figref\ref{sfig:entanglement_full}.
We reconstruct each Wigner function individually to produce the density matrix of cavity 1, conditioned on the measurement outcome in cavity 2.
As before, the measured Wigner function is normalized before reconstruction, to correct for measurement contrast in the parity mapping and ancilla readout.
The result is two conditional density matrices for each of the three bases.
This way we can reconstruct the two-qubit state without having to apply a decode pulse on cavity 1 as well.

We use these conditional density matrices to reconstruct the logical two-qubit density matrix.
This reconstruction uses a routine used in \refref\cite{Chou2018}, adapted for our tomography scheme.
The use of the $\ket{f}$ level of the ancilla for enhanced and roughly symmetric measurement contrast obviates the need for additional measurements to symmetrize the resultant data (see \secref\ref{sssec:tomo_readout}).

Each joint choice of bases for the two cavities is given by $\{k,l\}\in \{x,y,z\}^{\otimes 2}$, where $k$ corresponds to the basis choice for cavity 1 and $l$ for cavity 2.
We refer to the measured probabilities of the logical measurements in cavity 2 as $p_{\pm l}$, and the conditional density matrices of cavity 1 as $\rho_{\pm l}$.
The goal is to produce the expectation values $p_{\pm k,\pm l}$ of the four projectors $\Pi_{\pm k,\pm l} = \ket{\pm k}_1 \ket{\pm l}_2 \bra{\pm k}_1 \bra{\pm l}_2 $, which correspond to the probability of measuring the joint state to be in $\ket{\pm k}_1 \ket{\pm l}_2$..
This joint probability is $p_{\pm k,\pm l} = p_{\pm l} P( \pm k | \pm l)$, where $P( \pm k | \pm l)$ is the conditional probability of measuring $\pm k$ in cavity 1 given the result $\pm l$ in cavity 2.
This conditional probability is the expectation value of the single-cavity projector $\Pi_{\pm  k} = \ket{\pm k}_1 \bra{\pm k}_1 $, given the result $\pm l$ in cavity 2.

To compute these conditional probabilities, we take the conditional density matrix $\rho_{\pm l}$ and  evaluate $P( \pm k | \pm l) = \ev{\Pi_{\pm  k}}_{\pm l} \equiv \Tr{\rho_{\pm l} \Pi_{\pm  k}}$, which is the squared overlap of the measured cavity state with the logical state $\ket{\pm k}_1$ given outcome $\pm l$.
This is essentially the probability we would measure cavity 1 to be in $\ket{\pm k}_1$ with an ideal projective measurement.
It is important to note here that, since the cavity density matrix is of dimension larger than 2, leakage out of the logical subspace (here, $\{\ket{0},\ket{1}\}$) results in $P( + k | \pm l) + P( - k | \pm l) < 1$.
We will see in a moment that this results in a reconstructed logical two-qubit state with trace slightly less than 1.
Since we assign a binary outcome to the logical measurement of cavity 2, $p_{+l} + p_{-l} = 1$ by construction.
This means that leakage out of the logical space on cavity 2 is not directly observed.
However, such leakage will contribute to infidelity.
Since the decode operation cannot account for this leakage, the result is some arbitrary outcome of the ancilla measurement, which is we assume to be uncorrelated with the result in cavity 1.
Thus, while the decode-and-measure sequence will mask this leakage, it should convert it to infidelity in the form of a statistical mixture.
Put another way, this local operation cannot increase the amount of entanglement, so it does not result in overestimation of the fidelity.

The $3\times 3\times4=24$ computed joint probabilities $p_{\pm k,\pm l}$ are fed into the MLE reconstruction routine, which is a convex optimization over the space of all physical two-qubit ($2^2$ dimensional) density matrices.
To ensure physicality, the resultant density matrix $\rhoL$ is constrained to be Hermitian and positive semi-definite.
In addition, $\Tr{\rhoL} \le 1$ to account for the possibility of leakage out of the logical space as discussed above.
This leakage is very small for the $\{ \ket{0},\ket{1} \}$ encoding  \md the trace of $\rhoL$ (the value of the II bar) is found to be $0.999$, consistent with very small ($<10^{-3}$) occupation of Fock states above $n=1$ for the reconstructed cavity density matrices.

For the two-photon entangled state, there is a small but measurable amount of leakage outside of the $\{ \ket{0},\ket{2} \}$ code space due to errors in the parity measurement and cavity decay during tomography.
The value of the II bar (and hence the trace of $\rhoL$) is 0.991, consistent with a typical 1\% occupation of the $\ket{1}$ state in the measured Wigner functions of cavity 1.
In fact, the occupation of the error state $\ket{1}$ is found to be largest for states with large occupation of $\ket{2}$, suggesting cavity decay errors are primarily responsible.

\section{Cat code construction}\label{sec:cat_code}

\subsection{Definition}\label{ssec:cat_code_def}
Our error-correctable encoding is the four-component cat code \cite{Leghtas2013}.
The codewords have definite photon number modulo 4:
\begin{equation}
\begin{aligned}
\ketL{0} &= \frac{1}{\sqrt{N_0}}\left( \ket{\alpha}-\ket{i\alpha}+\ket{-\alpha}-\ket{-i\alpha} \right) \propto \sum_{n=2,6,10,...} \frac{\alpha^n}{\sqrt{n}!}\ket{n}\\
\ketL{1} &= \frac{1}{\sqrt{N_1}}\left( \ket{\alpha}+\ket{i\alpha}+\ket{-\alpha}+\ket{-i\alpha} \right) \propto \sum_{n=0,4,8,...} \frac{\alpha^n}{\sqrt{n}!}\ket{n}\\
\end{aligned}
\end{equation}
where $N_i$ denotes the state-dependent normalization factor.
These states are orthogonal for all values of $\alpha$.

A single photon loss event on the logical space spanned by these codewords takes a superposition of codewords into the error space which is also spanned by the odd-parity four-component cats:
\begin{equation}
\begin{aligned}
\ketE{0} &= \frac{1}{\sqrt{N_0 '}}\left( \ket{\alpha}-i\ket{i\alpha}-\ket{-\alpha}+i\ket{-i\alpha} \right) \propto \sum_{n=1,5,9,...} \frac{\alpha^n}{\sqrt{n}!}\ket{n}\\
\ketE{1} &= \frac{1}{\sqrt{N_1 '}}\left( \ket{\alpha}+i\ket{i\alpha}-\ket{-\alpha}-i\ket{-i\alpha} \right) \propto \sum_{n=3,7,11,...} \frac{\alpha^n}{\sqrt{n}!}\ket{n}\\
\end{aligned}
\end{equation}
An odd parity outcome after the state transfer tells us we just need to relabel the codewords to that of the error space and we will have mostly preserved the quantum information.

\subsection{Optimum cat size}\label{ssec:cat_code_size}
The value of $\alpha$ for the cat code is something we can chose when we encode in the initial states with NOCPs.
We can see that in the limit $\alpha\rightarrow 0$, the codewords become $\ketL{0} =  \ket{2}$, $\ketL{1} = \ket{0}$.
This is not a good encoding for error detection since only the $\ketL{0}$ codeword can lose a photon. Upon measuring odd parity, the state is projected into $\ket{1}$, destroying any initial superposition.
We can detect errors, but cannot recover the information.
For small but finite values of $\alpha$, upon knowing that a photon was lost, the state will be polarized more towards the codeword that started with the larger number of photons resulting in a loss of fidelity.
Similarly, upon knowing that no photon was lost, the state is polarized more to the codeword with fewer photons.
This no-jump backaction results in a logical dephasing error.

\begin{figure}[t]
    \includegraphics{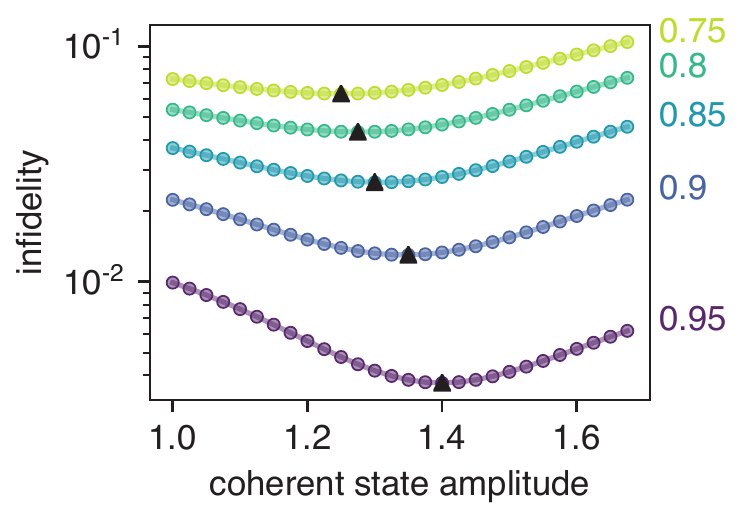}
    \caption{\label{sfig:fid_vs_alpha}
    \textbf{Optimum Cat Code Size.}
    For efficiencies ranging from 0.75 to 0.95 (text annotations at right), the ideal corrected infidelity versus cat size $\alpha$.
    Optimum fidelity denoted by black triangles.
    }
\end{figure}

This no-jump error is suppressed exponentially for large $\alpha$, since $\ketL{0}$ and $\ketL{1}$ will contain the same number of photons on average ($\bar{n} = |\alpha|^2$).
However, at larger $\alpha$, the dominant error is two-photon loss errors.
Since loss of two photons does not change the parity, this error is undetectable, and results in a logical bit-flip.
As such, there is an optimum value of $\alpha$ to use for a given energy transfer efficiency $\eta$ as illustrated in \figref\ref{sfig:fid_vs_alpha}.
This trade-off has been explored theoretically in more detail in \refref\cite{Li2017}.

In this experiment, $\eta = 0.84$ yields an optimal starting $\alpha = 1.3$.
With only photon loss error, this puts a theoretical upper bound of 0.97(1) for the transfer fidelity.
The measured value of 0.92 is lower due to additional experimental errors, mainly excitations of the conversion transmons and infidelity in the parity measurement and state preparation.

\subsection{Comparison to Binomial encoding}
The optimal cat code basis is qualitatively very similar to the lowest order binomial encoding \cite{Michael2016} with codewords $\ketL{0} = (\ket{0}+\ket{4})/\sqrt{2}$, $\ketL{1} = \ket{2}$.
For the experimental transfer efficiency $\eta = 0.84$, the cat code is predicted to give slightly better transfer fidelity by a few percent due to lower overhead.
This owes to the fact $\bar{n}\approx1.7$ for the optimal cat code vs $\bar{n} = 2$ for the binomial code.

\begin{table*}[t]
    \centering
    \begin{tabular}{ p{0.2\textwidth} | p{0.15\textwidth} p{0.16\textwidth} p{0.16\textwidth}}
    \hline\hline
    Source & Inefficiency & Infidelity (Fock) & Infidelity (Cat)\\
    \hline
    Bus loss                &  0.11  &  0.04  & 0.01 \\
    Transmon excitation     &  0.04  &  0.02  & 0.03 \\
    State preparation       &  0.01  &  0.01  & 0.02 \\
    Parity measurement      &  n/a   &  n/a   & 0.02 \\
    \hline
    Total                   &  0.16  &  0.07  & 0.07 \\
    Measured                & 0.16(1)& 0.08(1)& 0.08(2) \\
    \hline\hline
    \end{tabular}
    \caption{\label{tab:error_budget}
    \textbf{State transfer error contributions.}
    Error quoted due to bus loss for cat code is the remaining infidelity from second-order erorrs, assuming perfect error-tracking.
    ``State preparation'' error is fidelity of preparing a single photon for the inefficiency, and mean state fidelity of encoding for the infidelities.
    Infidelity due to parity measurement is roughly equal contributions of error in parity mapping and ancilla measurement, both of which can be suppressed by repeated fault-tolerant parity measurement \cite{Rosenblum2018}.
    }
\end{table*}

\subsection{Post-transfer basis}
After the transfer protocol, the information is encoded in a new logical basis.
After reconstructing the density matrix of the 6 cardinal transferred states, we find the basis that maximizes the average transfer fidelity independently for each parity outcome.
We optimize the choice of basis over the size of the cat, $\alpha$ and a deterministic phase shift that is different for the even and odd parity outcomes. For both parity outcomes, we find an optimal $\alpha = 1.2$, close to what we expect from the no-jump backaction.
Both these bases still have the same error-correctable properties as the original basis, namely further single photon loss events can be detected by measuring parity jumps.

\section{Error-Corrected state transfer}
To demonstrate the feed-forward tools needed to actively correct for photon loss during the state transfer, we extend the error-tracked state transfer with a real-time conditional decoding procedure.
We apply a decoding NOCP on module 2 to transfer the state from cavity 2 to ancilla 2, ideally leaving cavity 2 in the vacuum state, analogous to the decoding used for the logical measurement used for entanglement tomography  in \ssecref\ref{ssec:entangled_tomo}.
Overall, this amounts to transferring the qubit state from ancilla 1 to ancilla 2 with the information encoded in the $\ket{g},\ket{e}$ states of ancilla 2 at the end of the protocol.
Ancilla measurement for qubit tomography is the same as the measurement described in \secref\ref{sssec:basis_msmt}.

Since the qubit is encoded in a different basis depending on the measured error syndrome outcome, we pre-load two decodings, one for even parity and one for odd.
The controller branches on the parity measurement outcome to use the correct decoding operation, similar to \refref \cite{Ofek2016}.
The NOCPs for the Fock and cat encodings introduce a small additional infidelity estimated to be around 1\textendash{2}\% due to ancilla decoherence and pulse errors.
Since the ancilla can be treated a two-level system, leakage errors out of the cavity code space are converted to errors on the ancilla of another type, which in general depends on the details of the leakage and the NOCP.
While the decoding process masks the form of these errors, it cannot in principle improve the fidelity beyond what was measured from direct Wigner tomography.

For both Fock and cat encodings, we find an ancilla-to-ancilla average state transfer fidelity of 0.91, indicating we are also at break-even for this extended version of our protocol.
Qubit tomography and individual transfer fidelities are shown in \figref\ref{sfig:decoded} and \tabref\ref{tab:fids}.
Alternatively, one may error correct the cavities by performing conditional NOCPs that map the appropriate post-transfer basis back to the original encoded basis.

The decoded state tomograms revel important essential features of error correction with the cat code.
Since the codewords used do not have the same average photon number, there is a noticeable polarization error towards the state $\ketL{0}$ in the case of a single photon loss, evident in \figref\ref{sfig:decoded} b).
In other words, when we detect a single photon error, we learn something about the logical state: we were more likely to have started in state $\ketL{0}$ (contains $n = 2,6,...$), the state with larger average photon number.
Similarly, there is also a smaller polarization towards state $\ketL{1}$ (contains $n=0,4,...$) in the event of no photon loss.
These opposite polarizations cancel out in the wighted deterministic state, resulting in a symmetric loss of contrast in the X and Y bases, which is a logical dephasing error.
Also apparent in the deterministic data is a symmetric decrease in the Z polarization due to bit-flip errors from multi-photon loss events.
Since we operate at the optimum point, these logical bit and phase flip errors are balanced, and the result is a uniformly-depolarizing error.
A different choice of $\alpha$ can bias the logical error channel.
This trade-off is explored more fully theoretically in \refref \cite{Li2017}.

\section{Effectively deterministic two-photon entanglement}

\begin{figure}[t]
    \includegraphics{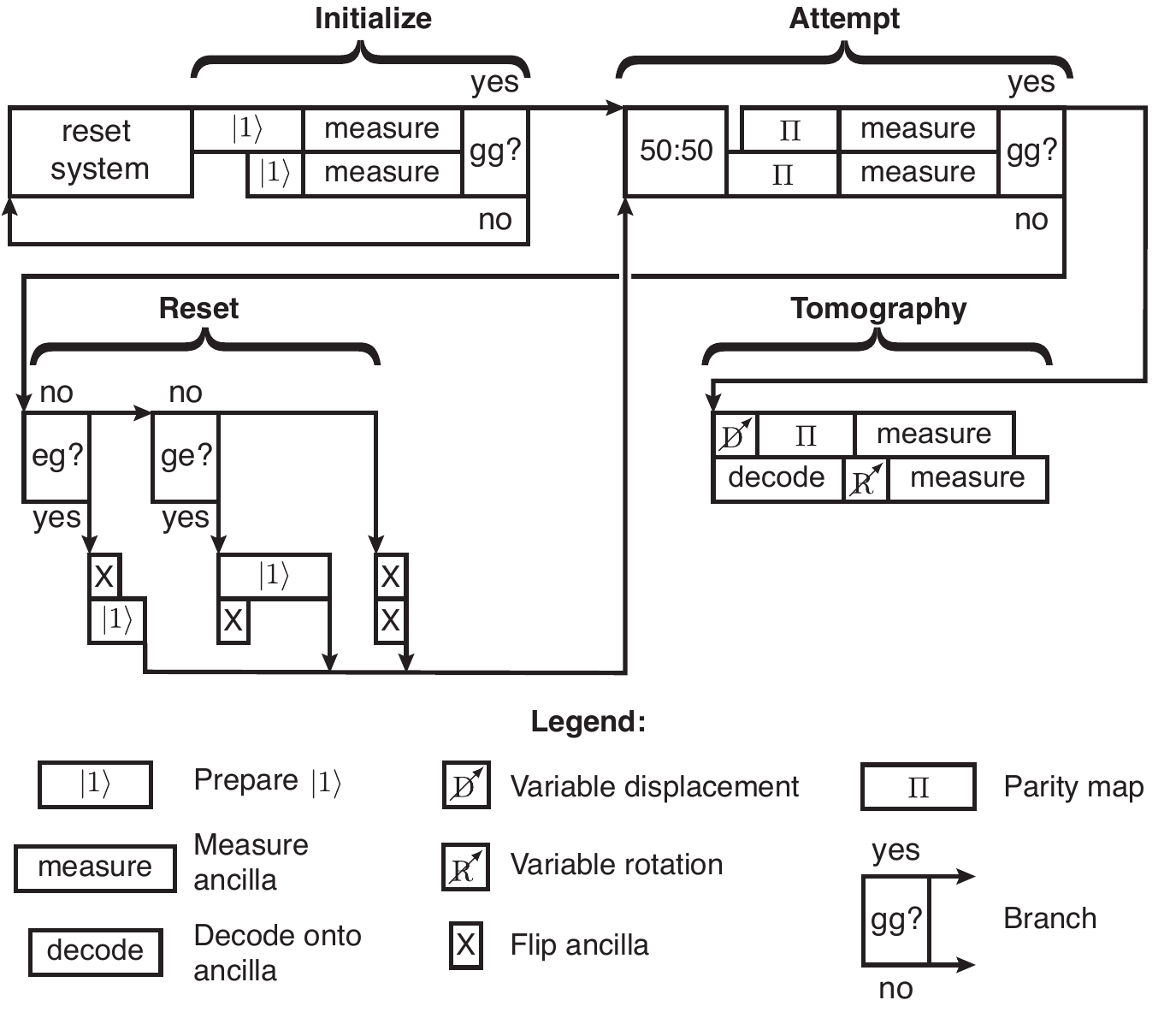}
    \caption{\label{sfig:repeated_hom_flow}
    \textbf{Repeated Hong\textendash{}Ou\textendash{}Mandel Entanglement Control Flow.}
    Top and bottom rows represented operations for modules 1 and 2, respectively.
    Boxes labeled with e.g. ``gg?'' represented binary conditional branch on ancilla measurements;   ``yes'' and ``no'' arrows are flow branch for measurement outcome.
    Box widths loosely represent operation times, but are not to scale.
    }
\end{figure}

For the Hong\textendash{}Ou\textendash{}Mandel entanglement scheme, in the event we measure a parity outcome other than (even,even), we in principal still know the current state of both cavities.
The parity outcomes in the event of single photon loss are (even,odd) and (odd,even) which occur with equal probability and project the cavities into the states $\ket{01}$ and $\ket{10}$.
We may reload a photon in the empty cavity using the corresponding NOCP to rapidly re-prepare the initial state $\ket{11}$ and try again to generate the desired Bell state within the same experimental realization.
We can repeat this protocol indefinitely until we obtain the desired (even,even) parity outcome, effectively making this scheme deterministic.
With this multi-round modification and keeping 100\% of the data, we can reach an average Bell state fidelity of 0.88(1), comparable to the single photon Bell state generation scheme.
If we impose a cutoff to the maximum number of rounds, we can boost this fidelity whilst maintaining a high success probability. This trade-off between Fidelity and maximum allowed number of rounds is shown in \figref\ref{sfig:hom_multiround}
Whilst this scheme mitigates the effect of single photon loss errors in the bus, errors from undesired transmon excitation become increasingly prevalent at high round number and result in failure to reload photons or enact the beam splitter, and inaccuracies in the parity measurements.
This is evidenced by the large number of rounds needed to reach failure probability near zero  in \figref\ref{sfig:hom_multiround}.
If the only error were photon loss in the bus, we would expect $>99\%$ success probability within three rounds.

The control flow for this repeated entanglement scheme is shown in \figref\ref{sfig:repeated_hom_flow}.
The flow is broken into several blocks: ``Initialize'' (2,208 \si{ns}, performed only once), ``Attempt'' (2,364 \si{ns}, repeated for each attempt), ``Reset'' (774 \si{ns} on average, repeated for each attempt after the first one), and ``Tomography'' (2280 \si{ns}, upon success).
Each readout block includes an acquisition delay for internal controller and cabling delays (360 \si{ns}), an acquisition time (580 \si{ns}), and a delay for the controller state estimation to be ready for branching (220 \si{ns}).
As explained in \ssecref\ref{sssec:tomo_readout}, the final tomography measurements have an additional ancilla $e-f$ rotation (40 \si{ns}), a longer acquisition (640 \si{ns}), and no controller delay.
The ``Reset'' length is not deterministic because of the differing lengths of the Fock state creation pulses and the differing number of decision branching steps (48 \si{ns}), so we quote the average.
The primary parity measurement outcomes are ``gg'' (success), ``eg,'' and ``ge'' (odd-even and even-odd, respectively), but there is a small ($\sim 1\%$) probability to measured ``ee'' (odd-odd) due to measurement errors.
In this case, we reset both ancilla and proceed as if the parity measurement were faithful, returning to the beamsplitter to try again.
We could instead measure the parity again to confirm or reject these outcomes as failures, but since these events are rare, the flow taken in this case is not very important.

Taking into account these sequence lengths and the relative probabilities of the number of rounds to success (see \figref\ref{sfig:hom_multiround}), the average time to success when considering up to three rounds is 5205 \si{ns}, as quoted in the main text, not including tomography.
The mean time to success for the fully-deterministic protocol (up to 147 rounds) is 6257 \si{ns}.
This time is only slightly longer because the probability of success approaches one in only a small number of rounds.

\section{Expanded data and sample parameters}

\subsection{Full state transfer data}
\begin{figure*}[ht]
    \includegraphics{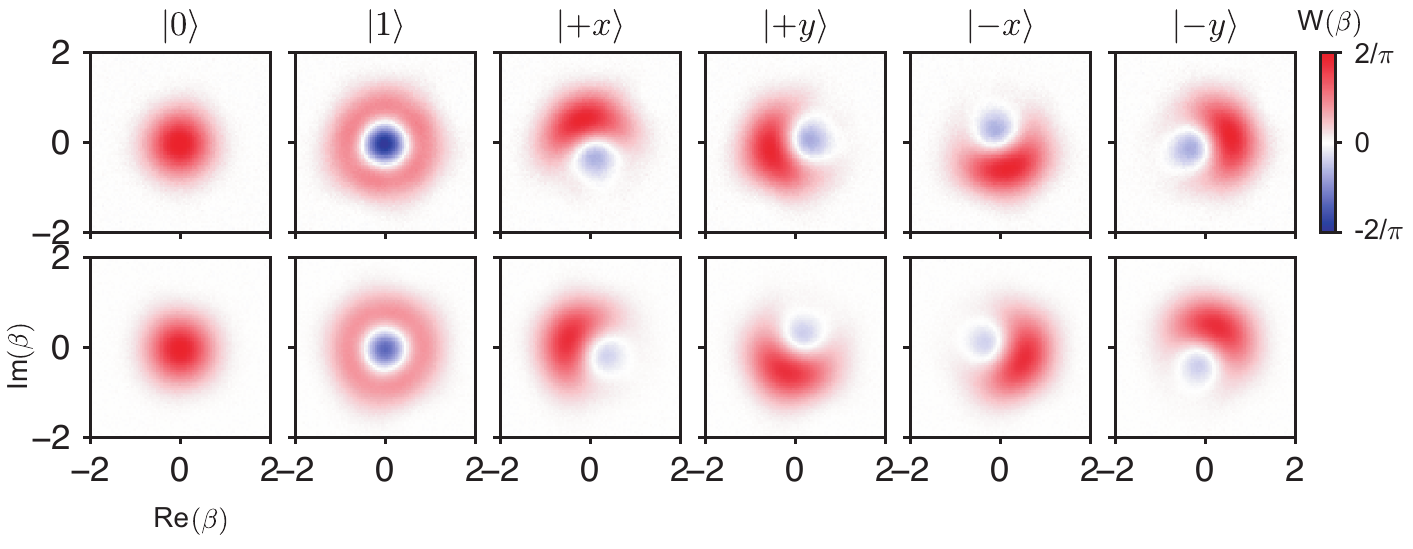}
    \caption{\label{sfig:fock_transfer_all}
    \textbf{Full Fock Code Data.}
    Measured Wigner functions for all six cardinal states of Fock encoding, as prepared in module 1 (top) and received in module 2 (bottom).
    }
\end{figure*}

\begin{figure*}
    \includegraphics{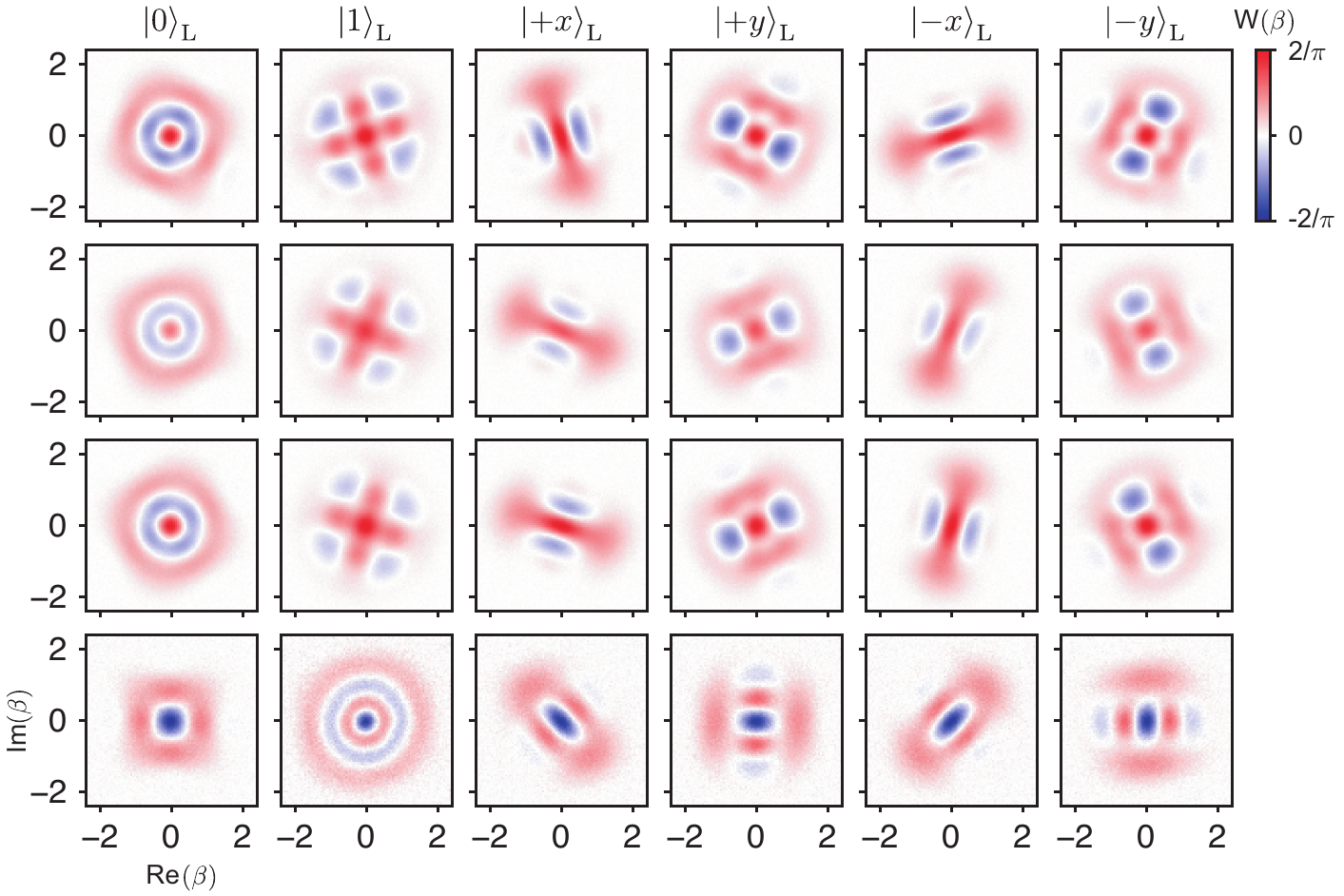}
    \caption{\label{sfig:cat_transfer_all}
    \textbf{Full Cat Code Data.}
    Measured Wigner functions for all six cardinal states of cat encoding, as prepared in module 1 (first row), received in module 2 with no parity measurement, (second row), and received and conditioned on even (third row) and odd (fourth row) parity.
    }
\end{figure*}

Wigner tomograms of all six states of the Fock encoding, measured as prepared in cavity 1 and received in cavity 2, are presented in \figref\ref{sfig:fock_transfer_all}.
Wigner tomograms of all six states of the Fock encoding, measured as prepared in cavity 1, received in cavity 2 without parity measurement, and sorted by parity after measurement, are presented in \figref\ref{sfig:cat_transfer_all}.
Decoded tomograms are shown in \figref\ref{sfig:decoded}.
State fidelities for all datasets are presented in \tabref\ref{tab:fids}.

\subsection{Full entanglement data}
\begin{figure*}
    \includegraphics{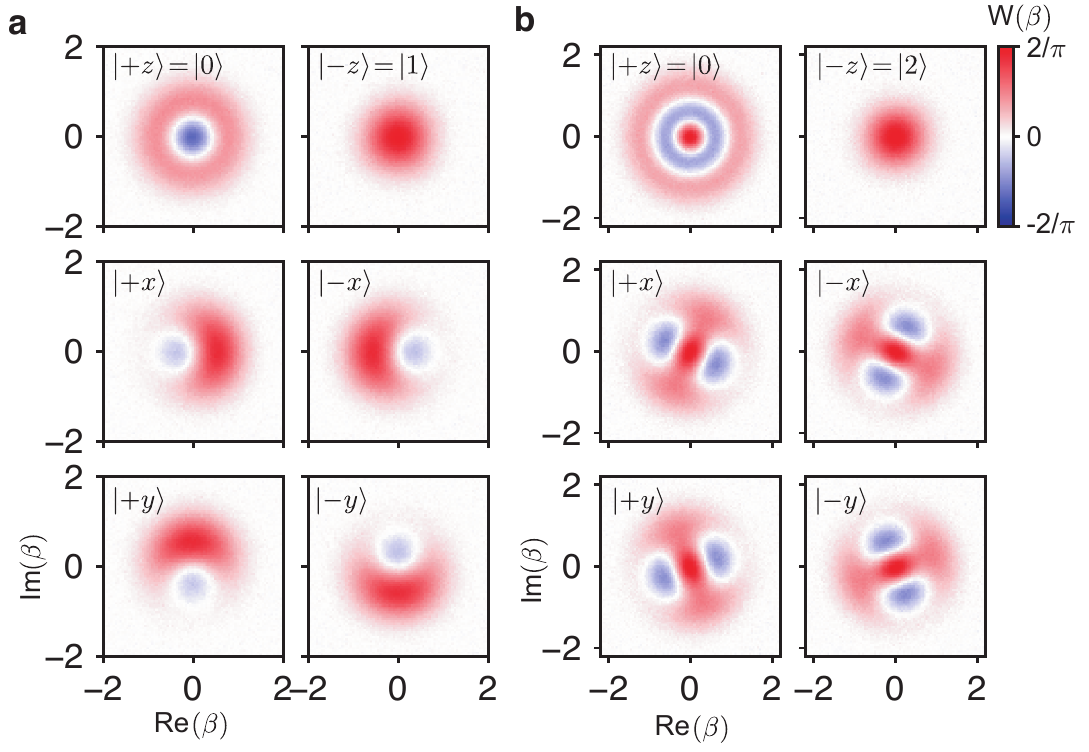}
    \caption{\label{sfig:entanglement_full}
    \textbf{Full Entanglement Data.}
    a) Single-photon entanglement tomography data for $z$, $x$, $y$ bases, top to bottom.
    b) Two-photon entanglement tomography data for $z$, $x$, $y$ bases, top to bottom.
    }
\end{figure*}

\begin{figure*}
    \includegraphics{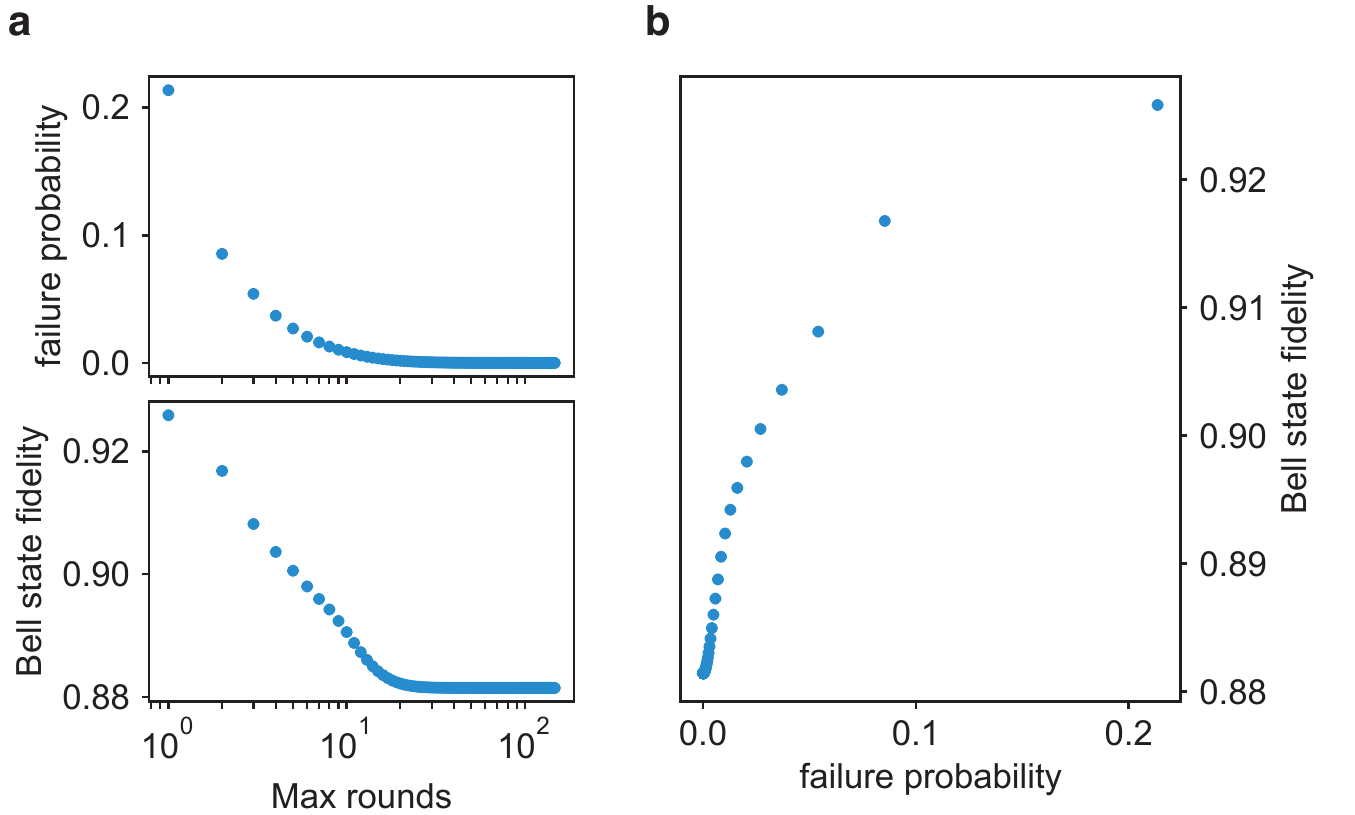}
    \caption{\label{sfig:hom_multiround}
    \textbf{Repeated Hong\textendash{}Ou\textendash{}Mandel Entanglement.}
    a) Cumulative failure probability and fidelity of repeated Hong\textendash{}Ou\textendash{}Mandel interference and parity measurement, as a function of the number of rounds accepted.
    b) Fidelity versus failure probability.
    }
\end{figure*}

\begin{figure*}
    \includegraphics[width=\textwidth]{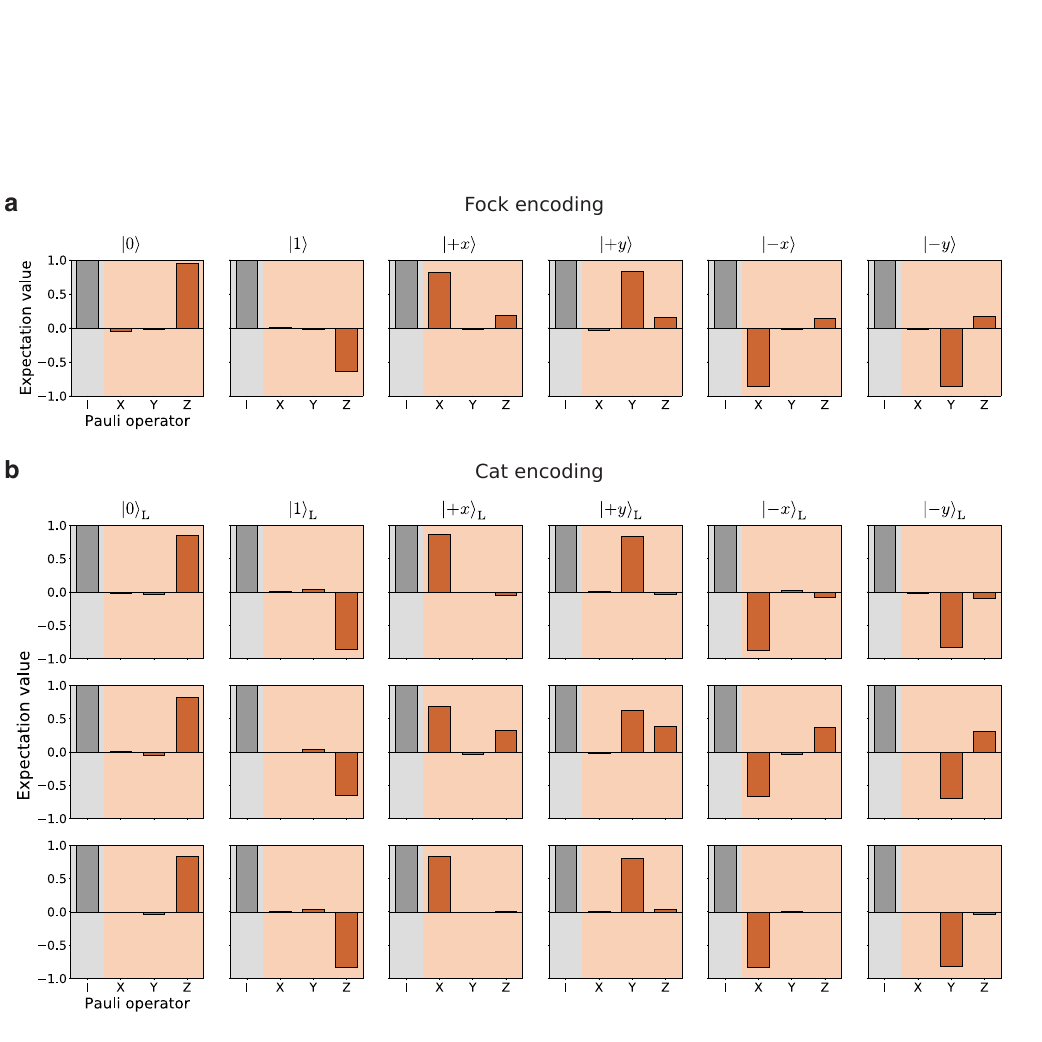}
    \caption{\label{sfig:decoded}
    \textbf{Decoded state transfer qubit tomography.}
    State transfer between ancilla 1 and ancilla 2 achieved with an additional decoding NOCP pulse on module 2.
    Qubit tomograms of ancilla 2 for the six cardinal states are shown for the two encodings.
    a) Using the Fock encoding for inter-module state transfer
    b) Using the cat encoding for inter-module state transfer, with parity-conditional decoding.
    Tomograms sorted by even parity outcomes (top row) and odd parity outcomes (middle), and full deterministic data (bottom),
    }
\end{figure*}

All six conditional Wigner tomograms for the single- and two-photon entangled states are presented in \figref\ref{sfig:entanglement_full}.
Results of multi-round two-photon entanglement are shown in \figref\ref{sfig:hom_multiround}.

\subsection{Sample parameters}
Measured sample parameters for both modules can be found in \tabref\ref{tab:devparams}.

\newpage

\begin{table*}[t]
    \centering
    \begin{tabular}{p{0.2\textwidth} p{0.12\textwidth} |p{0.13\textwidth} p{0.13\textwidth}}
    \hline\hline
    \multicolumn{2}{l}{Hamiltonian parameter (\si{\MHz}) }  & Module 1  & Module 2  \\
    \hline
    Mode frequency  & $\omega_\mathrm{a}/2\pi$      & $6514.3$               & $6505.2$  \\
                    & $\omega_\mathrm{t}/2\pi$      & $5838.5$               & $5668.7$   \\
                    & $\omega_\mathrm{c}/2\pi$      & $5081.6$               & $5149.2$   \\
                    & $\omega_\mathrm{r}/2\pi$      & $8970.7$               & $9014.9$  \\
                    & $\omega_\mathrm{f}/2\pi$      & $9077$                 & $9114$ \\
                    & $\omega_\mathrm{b}/2\pi$     & \multicolumn{2}{c}{5643\qquad\qquad\qquad} \\

    Pump frequency  & $\omega_\mathrm{X}/2\pi$      & $5222$              & $5228$ \\
                    & $\omega_\mathrm{Y}/2\pi$      & $6098$              & $6095$ \\

    Cross-Kerr      & $\chiat / 2\pi$       & $-1.138$                   & $-0.953$ \\
                    & $\chiac / 2\pi$       & $-0.765$                  & $-1.077$ \\
                    & $\chibc / 2\pi$       & $-4.3$                    & $-2.7$ \\
                    & $\chirt / 2\pi$       & $-1$                      & $-1$ \\

    Self-Kerr       & $\chiaa / 2\pi$       & $-3.9 \times 10^{-3}$      & $-4.1 \times 10^{-3}$ \\
                    & $\chitt / 2\pi$       & $-213.5$                 & $-203.4$ \\
                    & $\chicc / 2\pi$       &    $\dag$                 & $-112$ \\

    \hline
    \multicolumn{2}{l}{Decay parameter (\si{\us}) }   &  &  \\
    \hline
    Energy decay time       & $T^\mathrm{a}_{1}$ & $300$        & $450$  \\
                            & $T^\mathrm{t}_{1}$ & $35$         & $65$  \\
                            & $T^\mathrm{c}_{1}$ & $10$         & $20$  \\
                            & $T^\mathrm{r}_{1}$ &  $0.10$      &   $0.10$ \\
                            & $T^\mathrm{f}_{1}$ &  $0.004$     &   $0.005$ \\
                            & $T^\mathrm{b}_{1}$ & \multicolumn{2}{c}{$1.6$\qquad\qquad\qquad} \\

    Ramsey decay time       & $T^\mathrm{a}_{2\mathrm{R}}$        & $100$           & $140$ \\
                            & $T^\mathrm{t}_{2\mathrm{R}}$        & $15^*$          & $30$ \\
                            & $T^\mathrm{c}_{2\mathrm{R}}$        & $10$            & $20$ \\
    Hahn echo decay time    & $T^\mathrm{t}_{2\mathrm{E}}$        & $35$            & $80$ \\
                            & $T^\mathrm{c}_{2\mathrm{E}}$        & $20$            & $40$ \\

    \hline
    \multicolumn{2}{l}{Steady-state excitation } &  & \\
    \hline
    Ancilla                 & $1 - P(g)$        & $0.10$      & $0.12$ \\
    Cavity                  & $\bar{n}$         & $0.01$      & $0.01$ \\
    Converter               & $1 - P(g)$        & $0.1$      & $0.1$ \\
    \hline\hline
    \end{tabular}

    \caption[Sample Parameters]{\label{tab:devparams}
    \textbf{Sample Parameters.}
    Uncertainties of measured Hamiltonian parameters are $< 0.1\%$ except when indicated by fewer significant digits.
    Subscript f refers to Purcell filter.
    Decay parameters are observed to fluctuate around $10\%$; typical values are given.\\
    $\dag$ $\chicc$ of module 1 was not measured, but is expected to be similar to that of module 2.\\
    $*$ Ramsey decay of ancilla 1 was not a simple exponential, indicating a frequency instability.
    Reported number is decay scale at short times.
    }
\end{table*}

\begin{table*}[t]
    \centering
    \begin{tabular}{p{0.1\textwidth} p{0.09\textwidth} p{0.14\textwidth} |p{0.08\textwidth} p{0.08\textwidth} p{0.08\textwidth} p{0.08\textwidth} p{0.08\textwidth} p{0.08\textwidth} |p{0.07\textwidth}}
    \hline\hline
    & & &\multicolumn{6}{c|}{Transferred state} & \\
    Metric & Method & Parity outcome \rule[-1.5ex]{0pt}{0ex}& $\ketL{0}$ & $\ketL{1}$  & $\ketL{+x}$ & $\ketL{+y}$ & $\ketL{-x}$ & $\ketL{-y}$ & Average\\
    \hline
    \multicolumn{3}{l|}{Fock encoding} & & & & & & & \\
    \hline
    Fidelity    & Wigner & -   & 0.986 & 0.829 & 0.928 & 0.931 & 0.925 & 0.929      &  0.921 \\
                & Decoded & -  & 0.977 & 0.821 & 0.911 & 0.932 & 0.913 & 0.927      &  0.913 \\
    \hline
    \multicolumn{3}{l|}{Cat encoding} & & & & & & &\\
    \hline
    Probability & &$p_\mathrm{odd}$ & 0.218 & 0.104 & 0.162 & 0.162 & 0.162 & 0.160     & 0.161\\
    Fidelity    & Wigner&No syndrome& 0.721 & 0.867 & 0.808 & 0.793 & 0.802 & 0.787     & 0.796 \\
                &       & Even      & 0.915 & 0.950 & 0.940 & 0.923 & 0.943 & 0.921     & 0.932\\
                &       & Odd       & 0.939 & 0.824 & 0.859 & 0.840 & 0.858 & 0.856     & 0.863\\
                &       & Weighted  & 0.921 & 0.937 & 0.927 & 0.910 & 0.930 & 0.911     & 0.922 \\
                &Decoded& Even      & 0.923 & 0.927 & 0.934 & 0.917 & 0.936 & 0.918     & 0.923\\
                &       & Odd       & 0.909 & 0.823 & 0.840 & 0.810 & 0.832 & 0.846     & 0.843 \\
                &       & Weighted  & 0.920 & 0.916 & 0.920 & 0.900 & 0.919 & 0.908     & 0.913 \\
    \hline\hline
    \end{tabular}
    \caption[fids]{\label{tab:fids}
    \textbf{State transfer fidelities}
    Transfer fidelities of the reconstructed states shown in \figsref\ref{sfig:fock_transfer_all}, \ref{sfig:cat_transfer_all}, and \figref\ref{sfig:decoded}.
    Fidelities obtained from Wigner function reconstruction and from decoding onto ancilla are presented.
    For the cat encoding, fidelities for each parity outcome, as well as the probability of measuring odd parity, are provided.
    Fidelity is given for for cat code without a syndrome measurement.
    Weighted fidelity is the deterministic average fidelity of that state, weighted by the probability of the parity outcome.
    }
\end{table*}

\end{document}